\documentclass[12pt]{article}

\usepackage{latexsym}

\font\tenmsbm=msbm10 scaled 1200
\font\sevenmsbm=msbm9
\newfam\msbmfam
\textfont\msbmfam=\tenmsbm \scriptfont\msbmfam=\sevenmsbm

\makeatletter
\@addtoreset{equation}{section}
\makeatother

\newcommand{\eref}[1]{(\ref{#1})}

\def\be{\begin{equation}}
\def\ee{\end{equation}}
\def\ba{\begin{eqnarray}}
\def\ea{\end{eqnarray}}
\def\bet{\begin{tabular}}
\def\eet{\end{tabular}}
\def\pa{\partial}

\def\nn{\nonumber}
\def\ve{\varepsilon}

\def\a{\alpha}
\def\b{\beta}
\def\D{\Delta}
\def\d{\delta}

\def\r{\rho}
\def\s{\sigma}
\def\m{\mu}
\def\n{\nu}
\def\ra{\rightarrow}
\def\vp{\varphi}

\catcode`@=11

\begin{document}

\begin{titlepage}

\begin{flushright}
Preprint DFPD 06/TH/02\\
February 2006\\
\end{flushright}

\vspace{2.5truecm}

\begin{center}

{\Large \bf  Variational principle and energy--momentum tensor}
\vskip0.3truecm
 {\Large \bf for relativistic Electrodynamics of
point charges}

\vspace{1.5cm}

K. Lechner\footnote{kurt.lechner@pd.infn.it} and  P.A.
Marchetti\footnote{pieralberto.marchetti@pd.infn.it}
\vspace{2cm}

 {
\it Dipartimento di Fisica, Universit\`a degli Studi di Padova,

\smallskip

and

\smallskip

Istituto Nazionale di Fisica Nucleare, Sezione di Padova,

Via F. Marzolo, 8, 35131 Padova, Italia
}

\vspace{2.5cm}

\begin{abstract}

We give a new representation as tempered distribution for the
energy--momentum tensor of a system of charged point--particles,
which is free from divergent self--interactions, manifestly
Lorentz--invariant and symmetric, and conserved. We present a
covariant action for this system, that gives rise to the known
Lorentz--Dirac equations for the particles and entails, via Noether
theorem, this energy--momentum tensor. Our action is obtained from
the standard action for classical Electrodynamics, by means of a new
Lorentz--invariant regularization procedure, followed by a
renormalization. The method introduced here extends naturally to 
charged $p$--branes and arbitrary dimensions.

\vspace{0.5cm}

\end{abstract}

\end{center}
\end{titlepage}

\newpage

\baselineskip 8 mm
%originale
% \baselineskip 6 mm

%%%%%%%%%%%%%%%%%%%%%%%%%%%%%%%%%%%%%%%%%%%%%%%%%%%%%%%%%%%%%%%%%%%
%%%%%%%%%%%%%%%%%%%%%%%%%%%%%%%%%%%%%%%%%%%%%%%%%%%%%%%%%%%%%%%%%%%

\section{Introduction and Summary}

The dynamics of a classical charged point--particle interacting with
an electromagnetic field is described by the Maxwell and Lorentz
equations,
 \ba
 \pa_\m F^{\m\n}&=&j^\n, \quad  F^{\m\n}=\pa^\m A^\n-\pa^\n A^\m, \label{maxwell}\\
 {dp^\m\over ds}&=& e\,F^{\m\n}(y(s))\,u_\n(s).\label{lorentz}
 \ea
We parametrize the worldline of the particle $y^\m(s)$ by the proper
time $s$, and define four--velocity and four--acceleration
respectively by,
$$
u^\m={dy^\m\over ds},\quad w^\m={du^\m\over ds}.
$$
The current is,
 \be\label{2.2}
 j^\m (x) = e \int u^\m(s)\,
\d^4 (x - y (s))\,ds.
 \ee
Maxwell's equations admit the general solution,
$$
F^{\m\n}=F^{\m\n}_{LW}+ F^{\m\n}_{in},
$$
where $F^{\m\n}_{in}$ is a free external field, and $F^{\m\n}_{LW}$
amounts to the retarded Lienard--Wiechert (LW) field.

When substituting the total field strength into the Lorentz equation
one faces the problem that the LW--field is infinite on the
particle's trajectory, and the quantity,
 \be\label{LWself}
F^{\m\n}_{LW}(y(s)),
 \ee
diverges. This infinite self--interaction is, of course, a
consequence of the point--like nature of the charged particle.

There are several methods of dealing with this divergence, all of
them leading to the same conclusions: the quantity \eref{LWself} has
a divergent part, which leads to an infinite classical
renormalization of the particle's mass, and a finite part that
amounts to,
 \be
 F^{\m\n}_{LW}(y(s))\Big|_{finite}=-{e\over 6\pi}\left(u^\m
{dw^\n\over ds}- u^\n {dw^\m\over ds}\right).
 \ee
Substituting this into the Lorentz--equation one obtains the
relativistic Lorentz--Dirac equation,
 \be \label{3.1}
 {dp^\m\over ds}= {e^2 \over 6 \pi} \left({dw^\m\over ds}+w^2\,u^\m\right)
 + e\, F^{\m\n}_{in}(y) \,u_\n,
 \ee
which has to be considered as the equation of motion for the charged
particle, replacing the original ill--defined Lorentz--equation
\eref{lorentz}. Here, and in the following, we indicate the square
of a four--vector with $w^2\equiv w^\m w_\m$. Strictly speaking this
equation can not be {\it derived} from the fundamental equations
\eref{maxwell} and \eref{lorentz} of classical Electrodynamics (ED),
but has to be postulated.

In its relativistic form  equation \eref{3.1} has been derived first
by Dirac \cite{dirac} in 1938, in an attempt to describe the motion
of a classical relativistic charged point--particle, taking into
account its self--interaction. In this seminal paper the equation
was derived using in an essential way the principle of
energy--momentum conservation. This conservation law represents
still the basic motivation for the equation itself.

Despite of the fact that \eref{3.1} is a well--defined equation and
ensures energy--momentum conservation, it exhibits several
unsatisfactory features, both physically and mathematically. The key
physical drawbacks are that the equation is of the third order in
the time--derivative, and that it admits a class of solutions, so
called runaway solutions -- see e.g. \cite{thirring} -- which lead
to exponentially growing four--velocities, even in absence of
external forces. When these solutions are eliminated imposing
suitable boundary conditions at infinity \cite{haag,roh}, it turns
out that the motions {\it allowed} by these conditions show up an
acausal behavior, in the sense of a preacceleration: the
acceleration is felt before the external force begins to act.
Actually, this preacceleration occurs on a time scale of order,
 $$
 \tau ={e^2\over 6\pi m}\sim 10^{-23} s,
 $$
which is of a factor 1/137 shorter than the time scale
$\tau_q=\hbar/m$, where, due to the Heisenberg uncertainty
principle, quantum effects become relevant and the classical theory
has to be abandoned. Hence, physically one can interpret this
acausal behavior as an inconsistency of classical ED of
point--particles, which is cured by the quantum formulation of the
theory, see e.g. \cite{moniz} \footnote{One can, actually, maintain
a second order Lorentz--equation which is free from singularities,
if one introduces a finite cutoff of physical origin on the charge
distribution of the would--be point--particle. This is the
philosophy pursued mainly in \cite{spohn,yaghj} and references
therein. The main unsatisfactory aspects of this approach are its
lack of universality, and the problems related with the
implementation of exact energy--momentum conservation. In the
present paper we follow Dirac's original proposal and insist on
point--like charges, and hence on equation \eref{3.1}.}.

Accepting \eref{3.1} as the correct equation of motion for the
particle, three questions arise automatically, to which we will give
new answers in this paper. The first question regards {\it local}
energy--momentum conservation, in terms of a {\it well--defined
conserved energy--momentum tensor}. The second regards the existence
of an action from which \eref{3.1} can be derived, and the third
question is if the energy--momentum tensor can be derived from this
action via Noether's theorem.

Regarding the first question we mention that, although relying on
conservation criterions, in his paper \cite{dirac} Dirac does not
construct an energy--momentum tensor, nor does he give a recipe to
compute the four--momentum enclosed in a volume $V$ containing the
particle. The main obstruction to such a construction arises from
the particular form of the naive energy--momentum tensor of the
electromagnetic field,
 \be
 \label{3.2} {\tau}^{\m\n}_{em}= F^{\m\rho}_{LW}\, F _{LW}{}_\rho{}^\n + {1\over
 4}\,\,
\eta^{\m\n} F^{\rho\sigma}_{LW}\,F_{LW\rho\sigma},
 \ee
where for simplicity we have set the external field to zero. If we
indicate with $ R$ the distance from the particle at a certain
instant, the fields $F^{\m\n}_{LW}$ exhibit near the particle the
standard Coulomb--like {\it integrable} $1/R^2$ singularities, and
hence ${\tau}^{\m\n}_{em}$ exhibits the {\it non--integrable}
$1/R^4$ singularities. As a consequence the momentum integrals
$P_V^\m=\int_V d^3x\, \tau ^{0\m}_{em}$ diverge. In a mathematical
language the problem can be stated as follows: while the components
of $F^{\m\n}_{LW}$ are tempered distributions, i.e. elements of
${\cal S}'\equiv {\cal S}'({\bf R^4})$, the components of
${\tau}^{\m\n}_{em}$, being products of the formers, are not.

To obtain a well--defined energy--momentum tensor one has to replace
$\tau^{\m\n}_{em}$ with a tensor $T^{\m\n}_{em}$ whose components
are elements of ${\cal S}'$ such that, a) $T^{\m\n}_{em}$ coincides
with $\tau ^{\m\n}_{em}$ in the complement of the particle's
worldline, and b) such that the resulting total energy--momentum
tensor is conserved, if \eref{3.1} holds. Such a tensor admits then
automatically finite four--momentum integrals. This program has been
accomplished only rather recently by Rowe in \cite{rowese}, relying
partially on \cite{prerowe}. His energy--momentum tensor is defined,
rather implicitly, as the distributional derivative of the sum of
other elements of ${\cal S}'$, and entails the additional drawback
that it is not {\it manifestly} symmetric and traceless. On the
other hand it can be shown that the resulting tensor is uniquely
determined by the requirements a) and b).

In this paper we present an alternative and simple representation of
this tensor, relying on a new Lorentz--invariant regularization
procedure, followed by a renormalization, which leads to a
manifestly symmetric, traceless and invariant tensor. In particular,
our procedure follows the physical intuition that the renormalized
tensor should differ from \eref{3.2} only on the particle's
trajectory: correspondingly in our renormalization scheme we
subtract only (divergent) terms which are supported on the
world--line, i.e. terms that are proportional to delta--functions
supported on the trajectory.

An alternative approach to four--momentum conservation, based on a
different regularization procedure, has been proposed in
\cite{marino}. This approach leads to finite four--momentum
integrals, but it does not allow to construct an energy--momentum
tensor. Moreover, it lacks manifest Lorentz invariance.

The second question opened by \eref{3.1} is if this equation can be
derived from an action principle. It is immediately seen that the
resulting equation can not be derived from a {\it canonical} action.
Indeed, such an action would be necessarily of the type,
$$
I=-\int \left(m+e \, A^\m_{in} \,u_\m+ e^2\,{\cal
L}_{self}\right)ds,
$$
where ${\cal L}_{self}$ is an invariant function of the kinematical
variables $y^\m$, $u^\m$, $w^\m$ etc, giving rise to the
self--interactions at the r.h.s. of \eref{3.1}. But for dimensional
reasons the unique term with the right dimensions would be the
invariant ${\cal L}_{self}\propto u_\m w^\m$, and this is zero.

Since there exists no canonical action leading to \eref{3.1}, a
variational principle giving rise to this equation bears necessarily
some unconventional features. The first proposal for an action was
made in \cite{rohrprl}, relying on the previous attempt of
\cite{dirac}, and it requires the introduction of two gauge fields:
a source--free field and a field generated by the particle via
retarded {\it and} advanced potentials. As additional -- and more
serious -- drawback this action contains non--local Coulomb
interactions between the particles. The more recent attempt in
\cite{marino} provides an action involving a single gauge field, but
employs a regularization which is not manifestly Lorentz--invariant.

In this paper we present an action which is obtained from the
standard action for ED, with a single gauge field, in a very simple
and natural manner: we introduce a manifestly Lorentz--invariant
regularization in the standard action, and renormalize it upon
subtracting an infinite mass term. The so obtained action gives rise
to \eref{3.1} in the following way. First one varies  the action
with respect to the gauge field, obtaining a regularized version of
the Maxwell equation. Then one varies the action with respect to the
coordinates of the particle and substitutes in the resulting
equation the solution of the regularized Maxwell equation. The limit
of this equation, when the regulator goes to zero, is \eref{3.1}.

The third question regards the derivation of the energy--momentum
tensor through Noether's theorem, from a covariant action. As far as
we know this problem has been addressed only in \cite{rowel}, using
a formalism introduced in \cite{zwanziger}. This paper is based on a
manifestly Poincar\`e--invariant, but rather unconventional
distribution--valued Lagrangian density. Its most unusual
characteristics are that the minimal interaction term $A^\m j_\m$ is
absent, and that the self--interaction of the particle is
represented by a total derivative. The most unsatisfactory feature
of the resulting action is, however, that it does not give rise to
the equations of motion, i.e. \eref{maxwell} and \eref{3.1}.
Eventually the role of the Lagrangian of \cite{rowel} is restricted
to give rise, through translation invariance and Noether's theorem,
to the conserved energy--momentum tensor given in \cite{rowese},
upon imposing the Lorentz--Dirac equation \eref{3.1} ``by hand".

The Lagrangian which we propose in the present paper, being in
particular manifestly Poincar\`e invariant, allows to apply
Noether's theorem in a canonical way, and the resulting
energy--momentum tensor arises precisely in the new form mentioned
above.

The formalism used in this paper relies on the power of
full--fledged distribution theory. We remember, indeed, that in the
case of point particles the energy--momentum tensor as well as the
Lagrangian density are necessarily distribution valued. The other
key ingredient is a manifestly Lorentz--invariant regularization
procedure. Within this approach, in summary, we can set up the
entire lagrangian formalism for the ED of point particles as
specified above: construction of a conserved and well--defined
energy--momentum tensor, construction of an action which gives to
the correct equations of motion, and implementation of the
Noether--theorem, based on this action.

The method developed in this paper is based on the invariant regularization
of the Green function for the D'Alambertian in eq. \eref{2.7}, amounting
to the replacement $x^2\rightarrow x^2-\ve^2$, that permits a clear and simple distinction
between finite and divergent terms in the energy--momentum tensor. Since the
Green function in arbitrary dimensions depends
always on $x^2$,  see \cite{courant}, our method is very promising for  
the construction of finite and conserved energy--momentum tensors for charged
systems, for which this tensor has never been constructed before, \cite{prep}. Examples include point
particles in a curved background \cite{dewitt}, systems of dyons \cite{rohrdy},
point particles in higher dimensions \cite{kazinski}, and extended objects in arbitrary dimensions
\cite{kazinski2}. The knowledge of such a tensor allows a systematic quantitative analysis of 
radiation effects in these systems.

The plan of the paper is the following. In section two we introduce
the regularization used in the paper. In section three we present
and illustrate our main results, i.e. the new expression of the
renormalized energy--momentum tensor together with its main
properties, the renormalized Lagrangian density giving rise to the
equations of motion, and the Noether--theorem. Section four is
devoted to the proof of the properties of the renormalized
energy--momentum tensor, and to a comparison of this tensor with the
one proposed by Rowe \cite{rowese}. The variational results are
proved in section five. Some technical details are relegated to
appendices.

For simplicity we present our results in detail in the case of a
single point particle and for a vanishing external field
$F^{\m\n}_{in}$. The corresponding results for the general case are
summarized in section \eref{generali}.

\vskip1truecm

\section{Regularization}

In  Lorentz--gauge the Maxwell equations amount to,
 \be \label{2.6}
\Box A^\m = j^\m, \quad \partial_\m A^\m =0,
 \ee
where for a point--particle with charge $e$ and worldline $y^\m(s)$,
the current is given in \eref{2.2}. The solution of these equations
can be obtained in terms of the retarded Green function for the
D'Alambertian, supported on the forward light cone,
\begin{equation}
  \label{2.1}
G(x) = {1\over 2\pi} H (x^0)\, \d (x^2) = {1\over 4 \pi  |\vec
x|}\,\d (x^0 - |\vec x|), \quad\quad \Box G(x)=\d^4(x),
 \end{equation}
where $x^\m = (x^0, \vec x)$ are the space--time coordinates, and
$H$ denotes the Heaviside step function. Omitting the external
potential $A^\m_{in}$ one gets as solution the retarded
Lienard--Wiechert potential,
 \be
 \label{2.4} A^\m_{LW} =G*j^\m=
\left.{e\over 4\pi} {u^\m (s) \over (x_\n - y_\n (s)) u^\n(s)}
\right|_{s=s(x)},
 \ee
where $*$ denotes convolution, and the scalar function $s(x)$ is
uniquely fixed by the retarded time condition,
 \be
\label{2.5} (x - y (s))^2=0, \quad x^0>y^0(s),
 \ee
consequence of the $\d$--function appearing in \eref{2.1}.

Along the worldline of the particle, i.e. for $x^\m =y^\m(\lambda)$
for some $\lambda$, the LW--potential $A^\m_{LW}(y(\lambda))$
diverges. In the present notation this follows from $s (y(\lambda))
= \lambda$. As a consequence also the LW field--strength,
 \be\label{LWfs}
F_{LW}^{\m\n}=\pa^\m A_{LW}^\n-\pa^\n A_{LW}^\m,
 \ee
diverges along the particle worldline, meaning that a charged
particles feels an infinite self--interaction.

To smooth these singularities, while maintaining the structure of
the Maxwell equations unaltered, we propose to regularize the
retarded Green function in a Lorentz--invariant way. Choosing  an
UV--regulator $\ve >0$, with the dimension of length, we introduce a
regularized Green function \cite{fried},
 \be
 \label{2.7}
 G_{\ve} (x) = {1\over 2\pi}
H(x^0) \d (x^2 - \ve^2) \equiv {1\over 2\pi}\, \d_+ (x^2 -
\ve^2)={1\over 4 \pi \sqrt{|\vec x|^2+\ve^2}}\,\, \d \left(x^0
-\sqrt{|\vec x|^2+\ve^2}\right),
 \ee
which is still an element of ${\cal S}^\prime$. The support of
$G_{\ve}$ is given by the positive--time sheet of the hyperboloid,
$$
x^2= \ve^2.
$$
This regularization preserves therefore also causality.

We can now define a  regularized LW potential,
 \be
 \label{2.8}
A^\m_{\ve} \equiv  G_{\ve} * j^\m =\left.{e\over 4\pi} {u^\m (s)
\over (x_\n - y_\n(s)) u^\n(s)} \right|_{s=s_\ve(x)},
 \ee
which differs from \eref{2.4} only through the fact that  $s(x)$ is
replaced by $s_{\ve}(x)$, solution of the regularized retarded--time
equation,
 \be
\label{2.15} (x - y (s))^2=\ve^2, \quad  x^0>y^0(s).
 \ee
The potential $A^\m_{\ve}$ is a regular field, belonging to
$C^\infty\equiv C^\infty ({\bf R^4})$, and in particular it is
regular on the support of $j^\m$, i.e. on the worldline of the
particle. Also the regularized field strength,
 \be\label{fsreg}
F_\ve^{\m\n}\equiv\pa^\m A_\ve^\n-\pa^\n A_\ve^\m,
 \ee
is an element of $C^\infty$, and hence everywhere regular. We
illustrate these properties giving the explicit formulae for a
static particle, at rest in $\vec x=0$,
 \be\label{static}
A^0_\ve=   {e\over 4\pi} {1\over (|\vec x|^2+\ve^2)^{1/2}},\quad
  \vec A_\ve=0,\quad
  \vec E_\ve={e\over 4\pi} {\vec x\over (|\vec x|^2+\ve^2)^{3/2}},
\quad \vec B_\ve =0,
 \ee
which are all regular in $\vec x=0$.

One of the advantages of this regularization is that it preserves
the Lorentz--gauge. Indeed, due to the properties of the convolution
and thanks to the fact that the current is conserved, the definition
\eref{2.8} gives directly,
$$
\pa_\m A_\ve^\m=0.
$$
We can then define also a {\it regularized and conserved} current
through,
 \be
\label{2.11} j_{\ve}^\n\equiv \pa_\m F^{\m\n}_\ve= \Box A^\n_{\ve},
\quad\quad \pa_\n j_{\ve}^\n=0,
 \ee
that belongs to $C^\infty$, too.

Since the derivative is a continuous operation in ${\cal S}^\prime$,
it is clear that we have the following limits,
 \ba
{\cal S}^\prime {\rm -} \lim_{\ve \ra 0} A^\m_{\ve} &=& A^\m_{LW},\\
{\cal S}^\prime {\rm -} \lim_{\ve \ra 0} F^{\m\n}_{\ve} &=&
F^{\m\n}_{LW},\\
{\cal S}^\prime {\rm -} \lim_{\ve \ra 0} j^\m_{\ve} &=&
j^\m,\label{limcur}
 \ea
where ${\cal S}^\prime {\rm-}\lim$ denotes the limit in the (weak)
topology of ${\cal S}^\prime$ \footnote{We remember that this means
that these limits hold on every test function $\vp\in {\cal
S}\equiv{\cal S}(\bf R^4)$,
$$
\lim_{\ve\ra0}A^\m_{\ve}(\vp) = A^\m_{LW}(\vp), \quad {\rm etc.}
$$}.
Moreover, since by construction $G_{\ve}$ is  Lorentz invariant,
$A^\m_{\ve}$,  $F^{\m\n}_{\ve}$, as well as $j^\m_{\ve}$ are Lorentz
covariant tensor fields.

We end this section writing more explicit expressions for
$F^{\m\n}_{\ve}$ and $j^\m_{\ve}$, since these will be used in the
following. For this purpose we define the vector fields,
 \be
 \label{2.17} R^\m (x)\equiv x^\m -
y^\m, \qquad \Delta^\m (x)\equiv (u R)\, w^\m - (w R)\, u^\m ,
 \ee
where the kinematical quantities $y$, $u$ and $w$ in these
expressions  are all evaluated at the the proper time $s_\ve(x)$,
determined from \eref{2.15}. In the following this functional
dependence in the regularized quantities will always be understood.
For the scalar product of two vectors we use the notation $(a b)=
a_\m b^\m$. In particular we have, see \eref{2.15},
 \be\label{r2}
R_\m R^\m=(R^0)^2-|\vec R|^2=\ve^2.
 \ee
The expression for the regularized LW field strength can be derived
in the same way as the known expression for the standard LW field,
\eref{LWfs}. To calculate it from \eref{2.8} one needs the
derivative of the function $s_\ve(x)$, that is obtained
differentiating \eref{r2},
$$
 {\partial s_{\ve} \over \partial x^\m} = {R_\m
\over (uR )}.
 $$
The rest of the calculations is a bit lengthy but straightforward.
With the above conventions the results read,
 \ba
A^\m_{\ve} &=& {e\over 4\pi} {u^\m \over (uR)},\label{potreg}\\
 F^{\m\n}_{\ve} &=& {e \over 4\pi} {1\over (uR)^3}
 \left(R^\m u^\n - R^\n u^\m + R^\m \Delta^\n - R^\n
 \Delta^\m\right),\label{2.18}\\
j^\m_\ve&=&\ve^2\,{e\over 4\pi}\left( {1\over
(uR)^4}\left[(uR)\,b^\m-(bR)\,u^\m\right]+{3(1-(wR))\over
(uR)^5}\left(u^\m+\Delta^\m\right)\right),\label{curreg}
 \ea
where we defined,
 \be\label{dw}
b^\m={dw^\m\over ds}.
 \ee
The formula \eref{2.18} coincides, actually, with the known
expression of the LW field strength $F^{\m\n}_{LW} $ -- see for
example \cite{roh} -- the unique difference being that the
kinematical quantities are evaluated at the retarded proper time
$s_\ve(x)$, rather then at $s(x)$.

In the following we will use an asymptotic condition on the
worldlines $y^\m(s)$ of the particle. We suppose that the particle's
acceleration is zero before a certain instant \footnote{ Actually,
the results we obtain in this paper are valid also if the
acceleration vanishes sufficiently fast for $s\ra -\infty$.},
 \be\label{asym}
w^\m(s)=0, \quad {\rm for} \,\, s<\bar s.
 \ee
Since we have in any case $\lim_{|\vec x|\ra \infty}s_\ve(x^0,\vec
x)=-\infty$, this asymptotic condition implies that at fixed time
the acceleration $w^\m(s_\ve(x))$ in \eref{2.18} vanishes for
sufficiently large $|\vec x|$. The same holds for the corresponding
acceleration $w^\m(s(x))$ in the unregularized field
$F^{\m\n}_{WL}$. This implies that the regularized and unregularized
LW--fields have the same fixed--time Coulomb--like asymptotic
behavior for large $|\vec x|$,
 \be
 \label{asymlw}
F^{\m\n}_\ve \sim {1\over |\vec x|^2}, \quad F^{\m\n}_{LW} \sim
{1\over |\vec x|^2}.
 \ee
These behaviors, that are valid also if the acceleration vanishes
sufficiently fast in the remote past, will ensure in particular that
the total four--momentum of the electromagnetic field is finite.

\vskip1truecm

\section{Main results}

\subsection{The energy--momentum tensor}

We begin with the construction of a consistent energy--momentum
tensor, accepting as equation of motion \eref{3.1}.

As explained in the introduction the naive energy--momentum tensor
$\tau_{em}^{\m\n}$ in \eref{3.2} is not an element of ${\cal S}'$ --
although $F^{\m\n}_{LW}$ is -- due its $1/|\vec R|^4$ behavior near
the worldline \footnote{One can evaluate the quantity $|\vec
R(t,\vec x)|=|\vec x-\vec y(s(x))|$ at a point near the worldline,
$\vec x\approx \vec y(t)$, solving \eref{2.5}. One obtains,
$$
|\vec R(t,\vec x)|=\left({v \cos \a + \sqrt{1-v^2 \sin^2 \a}\over
1-v^2}\right) |\vec x-\vec y(t)|+o\left(|\vec x-\vec y(t)|^2\right),
$$
where $\vec v$ is the velocity at $t$, and $\a$ is the angle between
$\vec v$ and $\vec x-\vec y(t)$. Therefore, near the worldline
$|\vec R|$ represents indeed the spatial distance from the particle,
a part from a never vanishing constant.}. Therefore its derivatives
do not even make sense, and the question what the quantity
$``\pa_\m\tau_{em}^{\m\n}"$ amounts to, is meaningless.

To construct an energy--momentum tensor that is an element of ${\cal
S}'$ we start from the regularized $LW$ field \eref{2.18}, and
define the regularized electromagnetic energy--momentum tensor as,
 \be
 \label{treg}
T^{\m\n}_\ve= F^{\m\rho}_\ve F_{\ve\rho}{}^\n + {1\over 4}\,
\eta^{\m\n}\, F_\ve^{\rho\sigma} F_{\ve\,\rho\sigma}.
 \ee
This tensor belongs to $C^\infty$ and shares with $\tau_{em}^{\m\n}$
the asymptotic behavior for large $|\vec x|$ at fixed time, implied
by \eref{asymlw},
 \be
 \label{temasym}
T^{\m\n}_\ve \sim {1\over |\vec x|^4}, \quad \tau_{em}^{\m\n}\sim
{1\over |\vec x|^4}.
 \ee
Moreover, away from the worldline, i.e. for $\vec x\ne \vec y(t)$,
we have the {\it pointwise} limit,
 \be\label{point}
 \lim_{\ve\ra 0}
T^{\m\n}_\ve(x)= \tau_{em}^{\m\n}(x).
 \ee
But this limit does not exist in the topology of ${\cal S}'$, due to
the {\it reemerging singularities of $\tau^{\m\n}_{em}$ along the
worldline}. Before taking the ${\cal S}'$--limit one must isolate,
and subtract, these singular terms.

In the present framework the form of these singular terms -- and
this is one more advantage of our regularization -- is extremely
simple. Since the singularities appear only along the wordline, the
counterterm must be {\it covariant, symmetric, traceless and
supported along the wordline}. These requirements fix it essentially
uniquely, a part from an overall constant.  We propose as
renormalized energy--momentum tensor for the electromagnetic field
generated by a point--particle, the expression,
 \be
\label{3.5} T^{\m\n}_{em} \equiv {\cal S}^\prime - \lim_{\ve\ra 0}
\left[T^{\m\n}_\ve - {e^2\over 32\,\ve} \int \left(u^\m u^\n -
{1\over 4} \,\eta^{\m\n}\right)\d^4 (x - y
(s))\,ds\right]\equiv{\cal S}^\prime - \lim_{\ve\ra 0} \widetilde
T^{\m\n}_\ve.
 \ee
The first counterterm, proportional to $u^\m u^\n$, can be viewed as an electromagnetic
mass renormalization and is well known, see \cite{teit2}. The
second counterterm, proportional to $\eta^{\m\n}$, to our knowledge has
never been noticed before in the literature, 
probably because it can not be interpreted as a mass term and amounts, therefore, to a type of 
singularity that is not present in the Lorentz--equation \footnote{The technical reason for the omission
of this term stems from the fact that in the literature the momentum conservation equation is
derived integrating the formal (divergent) energy--momentum tensor $\tau^{\m\n}_{em}$ over a {\it restricted} 
class of spacelike surfaces, that are orthogonal to the four velocity $u^\m$. In this 
way one looses a divergent term proportional to  
${1\over\ve} \int \left(u^\m u^\n - \,\eta^{\m\n}\right)\d^4 (x - y(s))\,ds$.}. This term is, however, necessarily present because the divergent part of a traceless tensor must be traceless. 

The properties of the tensor \eref{3.5} are indeed summarized by the following

{\bf Theorem I}
\newline
{\it 1) The limit in} \eref{3.5} {\it exists and hence
$T^{\m\n}_{em} \in {\cal S}'.$
\newline
2) $T^{\m\n}_{em}$ is a Lorentz--covariant, symmetric and traceless
tensor field.
\newline
3) $T^{\m\n}_{em} (x)= \tau^{\m\n}_{em}(x)$ for $x$ in the
complement of the particle's wordline.
\newline
4) The four--divergence of $T^{\m\n}_{em}$ amounts to,
 \be\label{divtem}
 \pa_\m T^{\mu\nu}_{em}=- {e^2\over
6\pi} \int \left({dw^\n\over ds} +w^2
u^\n\right)\delta^4(x-y(s))\,ds,
 \ee
for an arbitrary smooth worldline $y^\m(s)$.
\newline
5) If one defines the energy momentum tensor of the particle in a
standard way as,
 \be
 T^{\m\n}_p  = m \int u^\m u^\n \d^{4} (x
- y(s))ds, \label{3.3}
 \ee
then the total energy--momentum tensor of the ED of a
point--particle,
 \be
\label{3.4} T^{\m\n} = T^{\m\n}_{em} + T^{\m\n}_p,
 \ee
is conserved, $\pa_\m T^{\mu\nu}=0$, if the particle satisfies the
Lorentz--Dirac equation} \eref{3.1}, {\it with $F^{\m\n}_{in}=0$}.

The crucial points to prove are the properties 1) and 4), i.e. the
existence of the limit \eref{3.5} and the evaluation of $\pa_\m
T^{\mu\nu}_{em}$;  we relegate these proofs to section \ref{proof1}.

The limit in \eref{3.5} holds, actually, also in a stronger sense,
i.e. on a set of test functions that is larger then ${\cal S}$.
 Indeed, since for a fixed time $x^0=t$, for $\ve\ra 0$ the tensor
$T^{\m\n}_\ve(t,\vec x)$ develops only a singularity in three--space
at the point $\vec x=\vec y(t)$, and due to the asymptotic behavior
\eref{temasym}, the limit \eref{3.5} exists also on ``pseudo--test
functions" of the form,
 \be
 \label{larger}
\vp(x)= \delta(x^0-t)\,\vp(\vec x),
 \ee
where  $\vp(\vec x)$ is bounded in ${\bf R^3}$, and of class
$C^\infty$ in a neighborhood of $\vec y(t)$. For example, $\vp(\vec
x)$ can be a constant, or a characteristic function on a
three--volume $V$. We will take advantage from this fact when
considering momentum integrals.

Property 2) holds by construction, and property 3) follows from the
fact that the counterterm in \eref{3.5} is supported entirely on the
worldline. This property has its physical origin in the fact that
the form of $\tau^{\m\n}_{em}$ off the worldline is regular, and its
phenomenological consequences away from the particle, like all
classical radiation phenomena, are experimentally very well tested.
We were therefore only allowed to change the form of
$\tau^{\m\n}_{em}$ on the worldline. Property 5) follows from
property 4), using \eref{3.1}.

\subsubsection{Momentum integrals}

The main purpose of this subsection is to prove, using the above
theorem, that the total four--momentum of the electromagnetic field
generated by a point--particle at a generic instant $t$ is given by
formula \eref{ptot}, as shown first in \cite{rowese}. In the course
of the proof we will illustrate how the new representation
\eref{3.5} can be applied in practice.

 Formula \eref{3.5} entails indeed an operative definition for the
four--momentum of the electromagnetic field, $P^\m_V(t)$, contained
at the instant $x^0=t$  in a volume $V$. We define this momentum
applying $T^{\m 0}_{em}$ to the pseudo--test function,
 \be\label{pseudo}
\vp(x)=\d(x^0-t)\chi_V(\vec x),
 \ee
where  $\chi_V(\vec x)$ is the characteristic function on $V$. By
definition we have,
 \be
 \label{momint}
P^\m_V(t)\equiv T^{\m 0}_{em}(\vp)= \lim_{\ve\ra 0} \widetilde
T^{\m0}_\ve(\vp)=\lim_{\ve\ra 0}\int_V \widetilde T^{\m0}_\ve(t,\vec
x)\,d^3x.
 \ee
This four--momentum is well--defined whenever $\vec y(t)\notin \pa
V$.

If at the time $t$ the particle is outside $V$, the counterterm in
\eref{3.5} does not contribute, and thanks to \eref{point} the above
definition reduces to,
$$
P^\m_V(t)=\int_V \tau^{\m 0}_{em}(t,\vec x)\,d^3x,
$$
coinciding with the standard expression.

We illustrate the  operative definition \eref{momint}, computing the
four--momentum $P^\m_V(t)$ for a particle in uniform motion,  the
result being in particular useful for the derivation of \eref{ptot}.
Due to Lorentz--invariance it is sufficient to consider a static
particle in the origin. From \eref{momint} we see that we have two
equivalent ways to evaluate $P^\m_V(t)$: we can evaluate $T^{\m
0}_{em}(\vp)$ on a generic pseudo--test function $\vp$, and then set
$\vp(x)=\d(x^0-t)\chi_V(\vec x)$, or we can evaluate the integral
$\int_V \widetilde T^{\m0}_\ve(t,\vec x)\,d^3x$, and then take the
limit $\ve \ra 0$. For a static particle we are able to carry out
the first, more ambitious, procedure.

Consider first the energy density. Inserting \eref{static} in
\eref{treg} one obtains the regularized energy density,
$$
T^{00}_\ve={1\over 2} \left({e\over 4\pi}\right)^2{r^2\over
(r^2+\ve^2)^3}, \quad r\equiv |\vec x|,
$$
and the $00$ component of \eref{3.5} reduces to,
$$
T^{00}_{em} \equiv {\cal S}^\prime - \lim_{\ve\ra 0} \left( {1\over
2} \left({e\over 4\pi}\right)^2{r^2\over (r^2+\ve^2)^3} -{3\,
e^2\over 128\,\ve}\,\d^3(\vec x)\right).
$$
Applying to a generic pseudo--test function $\vp$ we get,
 \ba\nn
T^{00}_{em}(\vp)&=& \lim_{\ve \ra 0} \left[{1\over 2} \left({e\over
4\pi}\right)^2 \int  {r^2 \vp(x^0,\vec x)\over (r^2+\ve^2)^3}\,
\,d^4 x-{3\, e^2\over 128\,\ve}\,\int \vp(x^0,\vec 0)\, dx^0\right] \\
&=&\lim_{\ve \ra 0}{1\over 2} \left({e\over 4\pi}\right)^2 \int {r^2
(\vp(x^0,\vec x)-\vp(x^0,\vec 0)) \over (r^2+\ve^2)^3} \,d^4 x\nn\\
&=&{1\over 2}
 \left({e\over 4\pi}\right)^2 \int {\vp(x^0,\vec x)-\vp(x^0,\vec 0)
\over r^4}\,d^4x, \label{cond}
 \ea
where we used,
$$
 {1\over 2}
\left({e\over 4\pi}\right)^2 \int
 {r^2\over (r^2+\ve^2)^3}\,d^3x = {3\,e^2\over 128\,\ve}.
$$
The integral in \eref{cond} is conditionally convergent, meaning
that one has first to integrate over angles, and then over $|\vec
x|$ and $x^0$. Inserting \eref{pseudo} we get for the energy
contained in $V$ the $t$--independent result,
$$
P^0_V= T^{00}_{em}(\vp)= {1\over 2} \left({e\over 4\pi}\right)^2
\int {\chi_V(\vec x)-\chi_V(0) \over r^4}\,d^3x.
$$
If $V$ does not contain the origin, i.e. the particle, one has
$\chi_V(0)=0$ and this formula reduces to the standard electrostatic
energy. If, on the other hand, $V$ is a sphere with radius $\rho$
centered in the origin one has $\chi_V(0)=1$, and the energy in the
sphere amounts to $P^0_V=-e^2/8\pi \rho$. In particular the total
energy in whole space is zero.

The spatial momentum $P^i_V (t)$ for a static particle vanishes
trivially, since the $0i$ components of both terms in \eref{3.5} are
zero, even before taking $\ve\ra 0$. Thanks to Lorentz--invariance
this means that the {\it total four--momentum of the electromagnetic
field generated by a charged particle in uniform motion is zero},
 \be\label{uniform}
P^\mu(t) =0.
 \ee

We consider now the total four--momentum for a particle in arbitrary
motion,
$$
P^\m(t)=\lim_{\ve\ra 0} \int\widetilde T^{\m 0}_\ve(t,\vec x)\,
d^3x,
$$
where, we remember, the convergence of the integral for large $|\vec
x|$ is ensured by the asymptotic behavior \eref{temasym}. Since for
$s<\bar s$ the particle is in uniform motion, we have the additional
information that,
 \be\label{pasym}
P^\m(t)=0, \quad {\rm for}\quad t<\bar t,
 \ee
thanks to \eref{uniform}, where $\bar t=y^0(\bar s)$.

To derive an explicit expression for $P^\m(t)$ we use property 4) of
the Theorem I. Define the vector field,
 \be
\label{km}
 K^\m_\ve= \pa_\n \widetilde T^{\n \m}_\ve.
 \ee
Since the derivative is a continuous operation, \eref{3.5} and
\eref{divtem} imply,
 \be
\label{porco}
 {\cal S}'-\lim_{\ve\ra 0} K^\m_\ve=- {e^2\over 6\pi}
\int \left({dw^\m\over ds} +w^2 u^\m\right)\delta^4(x-y(s))\,ds,
 \ee
where, again, this limit holds also on the pseudo--test functions
\eref{larger}. Integrating \eref{km} over whole three--space we
have,
$$
\pa_0 \int \widetilde T^{0\mu}_\ve(t,\vec x)\,d^3x+ \int \pa_i
\widetilde T^{i\mu}_\ve(t,\vec x)\,d^3x=\int  K^\m_\ve(t,\vec x)
\,d^3x.
$$
Using the three--dimensional Gauss theorem and \eref{temasym}, the
second term on the l.h.s. is zero. Taking then the limit $\ve\ra 0$
and using \eref{porco} one obtains,
$$
{dP^\mu(s)\over ds}=-{e^2\over 6\pi}\left({dw^\m\over ds} +w^2
u^\m\right),
$$
where we used the variable $s$ instead of $t=y^0(s)$. Integrating
this expression from $-\infty$ to $s$, and using the asymptotic
relations \eref{asym} and \eref{pasym}, we get for the total
four--momentum of the electromagnetic field,
 \be\label{ptot}
 P^\mu(s)=-{e^2\over 6\pi}\left(w^\m(s)+\int_{-\infty}^s
 w^2(\lambda)\,
 u^\m(\lambda)\,d\lambda\right),
 \ee
reproducing the result of \cite{rowese}.

In appendix C we will actually prove that the more implicit
expression for $T^{\m\n}_{em}$ given by Rowe in \cite{rowese},
defines the same distribution as \eref{3.5}.

The formula \eref{ptot} is clearly in agreement with total
four--momentum conservation, $P^\m(s)+p^\m(s)=$ constant, see
\eref{3.1}.

\subsection{Covariant variational principle}

The purpose of Theorem I was the construction of a well--defined
energy--momentum tensor. We deduce now the equations of motion and
the so constructed energy--momentum tensor -- via Noether theorem --
from an action principle.

We propose the following regularized Lagrangian density ${\cal
L}_\ve \equiv{\cal L}_{\ve}(A,y)$,
 \be
 \label{3.8}
 {\cal L}_{\ve} = -{1\over 4} F_{\m\n} F^{\m\n} - j^\m_{\ve} A_\m -
m_\ve \int \d^{4} (x - y(s))\,ds,
 \ee
that is an element of ${\cal S}'$ if $A^\m$ is sufficiently regular.
This Lagrangian differs from the standard Lagrangian for ED by two
conceptually simple ingredients. First, the current $j^\m$ has been
replaced by its regularized counterpart $j^\m_\ve$, see \eref{2.11}
and \eref{curreg}. Second, we introduced a diverging counterterm for
the mass. One can indeed interpret,
 \be\label{bare}
m_\ve \equiv m - {3\pi^2\over 8 \ve}\left({e\over 4\pi}\right)^2,
 \ee
as the bare mass of the regularized theory, whereas $m$ is the
physical renormalized mass.

We define the regularized action associated to this Lagrangian in a
standard way,
 \be
 \label{action}
 I_\ve[A,y]=\int {\cal L}_{\ve} \, d^4 x=
 -\int \left({1\over 4} F_{\m\n} F^{\m\n}+ j^\m_{\ve}
 A_\m\right)d^4x-m_\ve \int ds.
 \ee
Notice that this action is gauge invariant, because the current
$j^\m_\ve$ is identically conserved. Moreover, it gives rise to the
equations of motion \eref{maxwell} and \eref{3.1} according to the
following

{\bf Theorem II}
\newline
{\it 1) The equations of motion for $A^\m$ obtained from  $I_\ve$
are the regularized Maxwell equations,
 \be\label{maxreg}
\pa_\m F^{\m\n}=j^\n_\ve.
 \ee
These equations entail as solution the regularized LW potential
$A_\ve^\m$ in \eref{potreg}.
\newline
2) Consider the equations of motion for $y^\m(s)$ derived from
$I_\ve$,
$$
L^\m_\ve[A,y](s)\equiv {\d I_\ve[A,y]\over \d y_\m(s)}=0.
$$
If one substitutes for $A^\m$ the solution $A^\m_\ve$ of
\eref{maxreg}, one has the point--wise limit,
 \be\label{limvar}
\lim_{\ve\ra 0}L^\m_\ve[A_\ve,y](s)= {dp^\m\over ds}-{e^2 \over 6
\pi} \left({dw^\m\over ds}+w^2\,u^\m\right),
 \ee
corresponding to the Lorentz--Dirac equation.}

Property 1) is obvious. The statement 2) means that one has first to
derive the equation of motion for $y^\m$ from $I_\ve$, then one must
substitute in this equation the regularized LW--field potential
$A^\m_\ve$, and eventually perform the limit $\ve\ra 0$. This is
{\it not} equivalent to substituting $A^\m_\ve$ in \eref{action} and
then deriving the equation of motion for $y^\m$ and taking the limit
$\ve\ra 0$. This second procedure would give rise to the equation of
motion ${dp^\m\over ds}=0$, see Theorem III. The reason for the
failure of this procedure is that there exists no {\it canonical}
action giving rise to \eref{3.1}, as explained in the introduction.

Since we are not interested in the explicit form of $L^\m_\ve$, to
prove property 2) it is sufficient to vary $I_\ve$ w.r.t. $y^\m$,
set $A^\m =A_\ve^\m$ and then take the limit $\ve\ra 0$. The
equation \eref{limvar} is therefore equivalent to the relation,
 \ba
\left. \lim_{\ve\ra0}\, \d I_\ve[A,y]\right|_{A=A_\ve}&=&
  \lim_{\ve\ra 0}\left(  -\d\left(m_\ve \int ds\right) - \int A_{\ve\mu}\,\d j^\m_\ve\,d^4x\right)\nn\\
&=& \int\left({dp^\m\over ds}-{e^2 \over 6 \pi} \left({dw^\m\over
ds}+w^2\,u^\m\right)\right)\d y_\m\,ds,\label{var}
 \ea
where $\d$ means variation w.r.t. $y^\m$. The proof of this relation
will be given in section \ref{proof2}.

The regularity properties of the regularized Lagrangian itself are
expressed by

{\bf Theorem III}
\newline
{\it On the solution $A^\m_\ve$ the regularized Lagrangian converges
for $\ve\ra 0$,
$$
{\cal S}'-\lim_{\ve \ra 0} {\cal L}_{\ve}(A_\ve,y)  \equiv {\cal L}
\in {\cal S}',
$$
where, for a generic $\vp\in {\cal S}$,
 \be\label{lagren}
{\cal L}(\vp)={1\over2 }\left({e\over 4\pi}\right)^2 \int
{\vp(x)-\vp(y(s(x)))\over (uR)^4}\,d^4x - m \int \vp(y(s))\,ds.
 \ee
In this expression the regulator has been everywhere removed.}

The first integral in \eref{lagren} is conditionally convergent, as
\eref{cond}. The regularized lagrangian admits therefore a finite
limit on the solutions of \eref{maxreg}. Notice in particular that
the divergent contribution of the bare mass in \eref{action}
canceled out. Along the solutions $A^\mu_\ve$, also the action
$I_\ve$ itself admits a finite limit. It is obtained evaluating the
limit Lagrangian \eref{lagren} on the pseudo--test function
$\vp(x)=1$,
$$
\lim_{\ve\ra 0}I_\ve[A_\ve,y] = -m\int ds.
$$
The proof of this theorem is given in section \ref{proof3}.

As last result we state the Noether theorem.

{\bf Theorem IV}
\newline
{\it The invariance of the distribution ${\cal L}_{\ve}$ under
space--time translations, i.e. the identity,
 \be\label{deltal}
\d_t {\cal L}_\ve\equiv  \d A^\m {\d {\cal L}_\ve\over \d A^\m}-\d
y^\m {\d {\cal L}_\ve\over \d y^\m}  -a^\m \pa_\m {\cal L}_\ve = 0,
 \ee
with $\d y^\m = a^\m, $ $\d A^\m = a^\n\pa_\n A^\m $, if evaluated
on the solutions of the equations of motion \eref{maxreg} and
\eref{3.1}, implies for $\ve\ra 0$ the conservation law $\pa_\m
T^{\m\n}=0$, where $T^{\m\n}$ is defined in \eref{3.5}, \eref{3.3}
and \eref{3.4}. More precisely, on the solution of the equations of
motion one has the relation,
 \be
{\cal S}'-\lim_{\ve\ra 0}\, \d_t {\cal L}_\ve =a_\n \,\pa_\m
T^{\m\n}.
 \ee
}

The significance of this theorem is clear. We want only specify the
following conceptual point. Since ${\cal L}_\ve$ is an element of
${\cal S}'$ the variation in \eref{deltal}  is defined in the
distributional sense. This means that, by definition, we have,
 \ba
(\d_t{\cal L}_\ve)(\vp)&=&\int {\cal L}_{\ve} (x + a) \vp
(x+a)\,d^4x
-\int {\cal L}_{\ve} (x) \vp(x)\,d^4x\nn\\
&=&a^\m\int {\cal L}_\ve\,\pa_\m \vp \,d^4x+\widehat{\d}\int{\cal
L}_\ve\, \vp\,d^4x,\label{defvar}
 \ea
where $\widehat{\d}$ indicates the variation of the integral w.r.t.
$A$, {\it minus} the variation of the integral w.r.t. $y$. This is
due to the fact that the {\it explicit} dependence on $x$ of ${\cal
L}_\ve$ is through the difference $x^\m-y^\m(s)$ \footnote{In the
present case the definition \eref{defvar} needs only to be applied
to the term multiplying $m_\ve$ in \eref{3.8} -- because all other
terms are regular distributions -- in which case \eref{defvar}
gives, translated back to symbolic notation,
$$
\d_t \left(\int \d^{4} (x - y(s))\,ds\right)= -a^\m\pa_\m \left(\int
\d^{4} (x - y(s))\,ds\right) -\int \d y^\m{\pa\over \pa y^\m(s)}\,
\d^{4} (x - y(s))\,ds,
$$
that vanishes identically. In practice we can then always use
directly \eref{deltal}.}. The Proof of Theorem IV will be given in
section \ref{proof4}.

\subsection{The general case}\label{generali}

The results presented so far generalize easily to a generic set of
charged point particles with masses $m_r$ and charges $e_r$, and to
the presence of an external field $F^{\m\n}_{in}$. In this case the
total regularized electromagnetic field is given by,
 \be\label{fgen}
F^{\m\n}_\ve= F^{\m\n}_{in}+\sum_r F^{\m\n}_{(r)\ve},
 \ee
where $F^{\m\n}_{(r)\ve}$ is the regularized LW field produced by
the $r$--th particle. The electromagnetic energy--momentum tensor,
generalizing naturally \eref{3.5}, becomes,
 \be\label{tnew}
 T^{\m\n}_{em} \equiv
{\cal S}^\prime - \lim_{\ve\ra 0} \left[T^{\m\n}_\ve - {1\over 32\,
\ve}\sum_r e_r^2 \int \left(u^\m_r u^\n_r - {1\over 4}
\,\eta^{\m\n}\right)\d^4 (x - y_r(s_r))\,ds_r\right],
 \ee
with $T^{\m\n}_\ve$ given by \eref{treg} and \eref{fgen}.  There are
no divergent counterterms due to the new interactions, because the
off--diagonal terms in $T^{\m\n}_\ve$ are well defined
distributions, with only $1/R^2$ singularities near the worldlines,
and because $F^{\m\n}_{in}$ is supposed to be regular.

The four--divergence of this tensor is,
 $$ \pa_\m
T^{\mu\nu}_{em}=-\sum_r e_r\int \left[{e_r\over
6\pi}\left({dw^\n_r\over ds_r} +w^2_r
u^\n_r\right)+\left(F^{\n\m}_{in}(y_r)+\sum_{s\neq r}
F^{\n\m}_{(s)}(y_r)\right)u_{r\m}\right]\delta^4(x-y_r)\,ds_r ,
 $$
where $F^{\m\n}_{(s)}$ is the unregularized LW field of the $s$--th
particle. This relation follows directly from \eref{divtem}, since
the off--diagonal terms in $T^{\m\n}_\ve$ are free from overlapping
divergences, and on them one can simply apply the Leibnitz rule to
evaluate the four--divergence. The total energy--momentum tensor
$T^{\m\n}_{em}+T^{\m\n}_p$, with $T^{\m\n}_p  =\sum_r m_r \int
u^\m_r u^\n_r \d^4 (x - y_r(s_r))ds_r$, is then conserved if the
particles satisfy the Lorentz--Dirac equations,
 \be\label{ldnew}
{dp_r^\m\over ds_r}= {e_r^2\over 6\pi}\left({dw^\m_r\over ds_r}
+w^2_r u^\m_r\right)+e_r \left(F^{\m\n}_{in}(y_r)+\sum_{s\neq r}
F^{\m\n}_{(s)}(y_r)\right)u_{r\n}.
 \ee

Eventually also the regularized Lagrangian density admits the
natural generalization,
$$
{\cal L}_{\ve} = -{1\over 4} F_{\m\n} F^{\m\n} - j^\m_{\ve} A_\m -
\sum_r\left( m_r - {3\,e_r^2\over 128 \,\ve} \right)\int \d^{4} (x -
y_r(s_r))\,ds_r,
$$
where $j^\m_{\ve}$ is the sum of the regularized currents of the
particles. This Lagrangian gives rise to the Lorentz--Dirac
equations \eref{ldnew}, and entails via Noether theorem the
conserved energy--momentum tensor $T^{\m\n}_{em}+T^{\m\n}_p$. Along
the equations of motion ${\cal L}_{\ve}$ admits again a finite limit
for $\ve\ra 0$ in the topology of ${\cal S}'$.

\vskip1truecm

\section{Energy--momentum tensor}\label{proof1}

This section is devoted mainly to the proof of  Theorem I. The
ingredients needed are the same as those needed in the proofs of the
other theorems. For this reason we present here the necessary
technical tools in some detail, while we will omit them in the
proofs for the other theorems.

\subsection{Existence of the tensor $T^{\m\n}_{em}$}\label{extensor}

We start with the proof of the existence of the limit \eref{3.5},
meaning that we have to isolate in $T^{\m\n}_{\ve}$ the terms that
diverge for $\ve\ra 0$, if applied to test functions $\vp\in{\cal
S}$. Inserting \eref{2.18} in \eref{treg} we can rewrite the
regularized energy--momentum tensor as sum of three terms,
characterized respectively by the inverse powers $1/R^4$, $1/R^3$
and $1/R^2$,
 \be \label{4.1}
 T^{\m\n}_{\ve} =  \sum_{i=1}^3T^{\m\n}_{\ve i},
 \ee
where we count the dimensionful regulator $\ve$ as a power of $R$,
since we have $R^2=\ve^2$, see \eref{r2}. The explicit expressions
are,
 \ba
T_{\ve1}^{\m\n} &=& \left({e\over 4\pi}\right)^2  {1\over (uR)^6}
\left(- R^\m R^\n + 2(uR)u^{(\m} R^{\n)}- \ve^2 u^\m u^\n
\right)-{\rm tr},\label{t1}
\\
T_{\ve2}^{\m\n} &= &\left({e\over 4\pi}\right)^2 {1\over (u R)^6}
\left(2(wR)R^\m R^\n   +2(uR)R^{(\m}\D^{\n)} -2\ve^2 u^{(\m}
\D^{\n)}\right)-{\rm tr},\label{t2}
\\
 T_{\ve3}^{\m\n} &=& \left({e\over 4\pi}\right)^2 {1\over (uR)^6} \left( -
 \Delta^2 R^\m R^\n - \ve^2 \Delta^\m \Delta^\n\right) -{\rm
tr},
 \label{t3}
 \ea
where $-{\rm tr}$ means the subtraction of the trace of the tensor,
and we defined symmetrization by $a^{(\mu} b^{\nu)}={1\over 2}(a^\m
b^\n+a^\n b^\m)$.

We will now show that the divergent parts of these tensors, for
$\ve\ra 0$ in the topology of ${\cal S}'$, amount to,
 \ba
\left.T_{\ve1}^{\m\n}\right|_{div}&=&\left({e\over 4\pi}\right)^2
\int\left( {\pi^2\over 2\ve}\left(u^\m u^\n-{1\over
4}\eta^{\m\n}\right) +{16\pi\over 3}\,\ln\ve\,u^{(\m}
w^{\n)}\right)\d^4(x-y(s))\,ds,\label{div1}
 \\
\left.T_{\ve2}^{\m\n}\right|_{div}&=&-\left({e\over 4\pi}\right)^2
{16\pi\over 3}\,\ln\ve\int u^{(\m}
w^{\n)}\d^4(x-y(s))\,ds,\label{div2}
 \\
\left.T_{\ve3}^{\m\n}\right|_{div}&=&0. \label{div3}
 \ea
Summing up one sees that the logarithmic divergences cancel, and
that the polar divergence is canceled by the counterterm in
\eref{3.5}.

To derive \eref{div1}--\eref{div3} one must apply $T^{\m\n}_{i\ve}$
to a test function $\vp(x)$, and analyze the limit $\ve\ra 0$. In
doing this we encounter the technical difficulty that the
kinematical variables $y$, $u$ and $w$ appearing in
$T^{\m\n}_{i\ve}$ depend on $x$ in a rather complicated way, since
they are evaluated at the proper time $s_{\ve} (x)$. To disentangle
this implicit dependence we use the following identity for a generic
function $f\in C^\infty ({\bf R})$,
 \be
f (s_{\ve} (x)) = \int \d (s - s_\ve (x)) f(s)\,ds = \int 2 \d_+ [(
x - y (s))^2- \ve^2]\,  u^\m(s)\,(x - y (s))_\m f (s)\, ds.
\label{4.3}
 \ee
When we apply such a function to a test function it is then also
convenient to perform in the resulting integral the shift,
$$
x^\m \longrightarrow x^\m + y^\m (s).
$$

In the following we make a systematic use of this strategy,
beginning with the proof of \eref{div1}. Applying \eref{t1} to a
test function $\varphi$ one obtains then,
 \be
T_{\ve1}^{\m\n}(\vp)= A^{\m\n}_\ve(\vp)+B^{\m\n}_\ve(\vp),
 \ee
where, setting momentarily $e/4\pi=1$,
 \ba
 A^{\m\n}_\ve(\vp)&=& \int ds \int d^4x\,{2\d_+(x^2-\ve^2)\over
(u x)^5}\left(- x^\m x^\n + 2(ux) u^{(\m} x^{\n)}-\ve^2 u^\m
u^\n-{\rm tr}\right)\vp(y (s)),\nn
 \\
 B^{\m\n}_\ve(\vp)&=& \int ds \int d^4x\,{2\d_+(x^2-\ve^2)\over
(u x)^5}\left(- x^\m x^\n + 2(ux) u^{(\m} x^{\n)}-\ve^2 u^\m
u^\n-{\rm tr}\right)\cdot\nn\\
&&\hskip4.2truecm \cdot(\vp(y (s) + x)-\vp(y(s)).\label{bmn}
 \ea
All kinematical quantities are now evaluated in the integration
variable $s$. We have divided $T_{\ve1}^{\m\n}(\vp)$ in two terms,
adding and subtracting the term with $\vp(y (s))$. This separation
is convenient because we will see that $A^{\m\n}_\ve(\vp)$ gives
rise to the polar divergence in \eref{div2}, while
$B^{\m\n}_\ve(\vp)$ gives rise to the logarithmic divergence.

Actually, $A^{\m\n}_\ve(\vp)$ can be evaluated exactly, it suffices
to know the invariant integral, see appendix A,
 \be\label{invint}
  \int d^4x\, {2\d_+(x^2-\ve^2)\over (ux)^5}\, x^\m x^\n=
{1\over \ve}\int d^4x {2\d_+(x^2-1)\over (ux)^5}\, x^\m x^\n=
{\pi^2\over 4\ve} \left(5u^\m u^\n-\eta^{\m\n}\right).
 \ee
This gives,
$$
A^{\m\n}_\ve(\vp)={\pi^2\over 2\ve}\int \left(u^\m u^\n-{1\over
4}\,\eta^{\m\n}\right)\vp(y(s))\,ds,
$$
amounting to the polar divergence in \eref{div1}.

Consider now $B^{\m\n}_\ve$. From a dimensional analysis one sees
that $B^{\m\n}_\ve(\vp)$ diverges logarithmically as $\ve\ra 0$,
i.e. as $\ln\ve$. To extract this divergence it is then sufficient
to evaluate the limit,
$$
\lim_{\ve \ra 0}\, \ve\, {d\over d\ve}\,B^{\m\n}_\ve(\vp).
$$
To compute $\ve\, {d\over d\ve}\,B^{\m\n}_\ve(\vp)$ it is convenient
to rescale first in \eref{bmn} the integration variable $x\ra \ve\,
x$, then apply $\ve\, {d\over d\ve}$, and then rescale  back $x\ra
x/\ve$. Taking then the limit $\ve\ra 0$ the term $\ve^2 u^\m u^\n$
drops out and one gets \footnote{The resulting expression can be
evaluated explicitly noting the integral, with $y\equiv y(s)$,
 \ba
&&\int d^4x\,{2\d_+(x^2)\over (u x)^5} \,x^\m x^\n \,[\vp(y)-\vp(y +
x)+x^\a\pa_\a\vp(y+x)]=
 -\int d\Omega\, {m^\m m^\n\over (um)^5}\,m^\a\pa_\a\vp(y)=\nn\\
&& {4\pi\over 3} \left(u^\m \pa^\n\vp(y)+u^\n \pa^\m\vp(y)
  +\eta^{\m\n}u^\a\pa_\a\vp(y)-6\,u^\m u^\n
u^\a\pa_\a\vp(y)\right),\nn
 \ea
where $\int d\Omega$ denotes the integral over the
three--dimensional solid angle, and $m^\m=(1,{\vec x\over|\vec
x|})$.},
 \ba
 \nn \lim_{\ve \ra 0}\, \ve\,
{d\over d\ve}\,B^{\m\n}_\ve(\vp) &=& \int ds \int
d^4x\,{2\d_+(x^2)\over (u x)^5}\left(- x^\m x^\n + 2(ux)
u^{(\m} x^{\n)}-{\rm tr}\right)\cdot\\
&&\hskip0truecm \cdot[\vp(y(s))-\vp(y (s) +
x)+x^\a\pa_\a\vp(y(s)+x)]\nn\\
&=&-{8\pi\over 3} \int u^\m u^\n\,
{d\vp(y(s))\over ds}\,ds\nn\\
&=&  {8\pi\over 3} \int (u^\m w^\n+u^\n w^\m)\, \vp(y(s))\,ds.\nn
 \ea
 This is equivalent to,
 $$
B^{\m\n}_\ve(\vp)={8\pi\over 3}\,\ln\ve \int (u^\m w^\n+u^\n w^\m)\,
\vp(y(s))\,ds +o(1),
 $$
where $o(1)$ means terms that are regular for $\ve \ra 0$. This
gives rise to the logarithmic divergence in \eref{div1}.

The proof of \eref{div2} proceeds similarly,
 $$
T^{\m\n}_{\ve2}(\vp)=\int ds \int d^4x{2\d_+(x^2-\ve^2)\over (u
x)^5}\left(2(wx)x^\m x^\n + 2(ux) x^{(\m} \D^{\n)}-2\ve^2 u^{(\m}
\D^{\n)}-{\rm tr}\right)\vp(y (s) + x),
 $$
where here  $\Delta^\m \equiv (u x)\, w^\m - (w x)\, u^\m$. This is
again logarithmically divergent and one obtains, operating as above
\footnote{This time one needs the integral, with $y=y(s)$,
 \ba\nn &&
\int d^4x{2\d_+(x^2)\over (u x)^5}(x^\m x^\n x^\r) x^\a\pa_\a \vp(y+
x)= -\int d\Omega {m^\m m^\n m^\r\over (um)^5}\,\vp(y)=
\\
&&{4\pi\over
3}\left(\eta^{\m\n}u^\r+\eta^{\n\r}u^\m+\eta^{\r\m}u^\n-6u^\m u^\n
u^\r\right)\vp(y),\nn
 \ea
where $\int d\Omega$ is the integral over angles and $m^\m=(1,{\vec
x\over |\vec x|})$.},
 \ba
 \nn
\lim_{\ve \ra 0}\, \ve\, {d\over d\ve}\, T^{\m\n}_{\ve2}(\vp)&=&
\int ds \int d^4x{2\d_+(x^2)\over (u x)^5}\left(2(wx)x^\m x^\n +
2(ux) x^{(\m} \D^{\n)}-{\rm tr}\right)x^\a\pa_\a \vp(y(s) + x)\\
&=&-{8\pi\over 3}\int (u^\m w^\n+u^\n w^\m)\, \vp(y(s))\,ds.\nn
 \ea
This proves \eref{div2}.

The tensor $T^{\m\n}_{\ve3}$ admits trivially a finite limit for
$\ve\ra 0$, since only $1/|\vec R|^2$ powers are present. This
concludes the proof of property 1) of the theorem.

\subsection{Conservation of the energy--momentum
tensor}\label{proofcons}

The conservation of the total energy--momentum tensor is reduced to
the proof of property 4), i.e. to the evaluation of $\pa_\m
T^{\m\n}_{em}$, see \eref{divtem}. Since the derivative is a
continuous operation in ${\cal S}'$, once we have established the
existence of the limit \eref{3.5}, the operator $\pa_\m$ can be
interchanged with the limit, and we get,
$$
\pa_\m T^{\m\n}_{em}= {\cal S}'-\lim_{\ve\ra0} \left(\pa_\m
T^{\m\n}_\ve - {e^2\over 32\,\ve} \int \left(w^\n - {1\over
4}\,\pa^\n \right)\d^4 (x - y (s))\,ds\right).
$$
Since also the regularized LW--field satisfies the Bianchi identity,
$$
\ve_{\m\n\r\s}\,\partial^\n F^{\rho\sigma}_{\ve} =0,
$$
one deduces the standard result,
 \be \label{4.9}
\partial_\m T^{\m\n}_\ve = -F^{\n\m}_\ve j_{\ve\m},
 \ee
and hence,
 \be\label{interm}
 \pa_\m T^{\m\n}_{em}={\cal S}'-\lim_{\ve\ra0}
\left(-F^{\n\m}_\ve j_{\ve\m}-{e^2\over 32\,\ve} \int \left(w^\n -
{1\over 4}\,\pa^\n \right)\d^4 (x - y (s))\,ds\right).
 \ee
The calculation is therefore reduced to the evaluation of
$F^{\n\m}_\ve j_{\ve\m}$ for $\ve\ra 0$, in the topology of ${\cal
S}'$.

 A somewhat delicate point in the evaluation of this limit
is the following. The formula for the current \eref{curreg} carries
a factor of $\ve^2$ in front, due to the fact that in the limit
$\ve\ra 0$ the current $j^\m_\ve(x)$ goes to zero for $x^\m$ in the
complement of the wordline. Actually, it can be seen that when
evaluating the limit \eref{limcur}, there is only one term in
\eref{curreg} that has a non--vanishing limit, more precisely,
$$
{\cal S}^\prime - \lim_{\ve \ra 0}\left( \ve^2 {e\over 4\pi}
{3u^\m\over (uR)^5}\right)=j^\mu,
$$
while all other terms converge to zero. These ``vanishing"
additional terms are however needed in the regularized current to
ensure its conservation, and they contribute in the product
$F^{\n\m}_\ve j_{\ve\m}$ for $\ve \ra 0$, since $F^{\n\m}_\ve$ has
polar terms. Multiplying out \eref{2.18} and \eref{curreg} we obtain
indeed,
 \ba
F^{\n\m}_{\ve} j_{\ve \m}&=& \left({e\over 4\pi}\right)^2 \ve^2
\left[
 \left({2w^2 \over (uR)^6} -{(b R)\over (uR)^7}\right)R^\n +3\,{(wR)-1\over
 (uR)^6}\, w^\n\right.\nn
\\
&& -\left.3\,{((wR)-1)^2 \over (uR)^7}\left(u^\n+((wR)-1){R^\n\over
(uR)}\right)\right] +o(\ve),
 \label{4.14}
 \ea
where  $b^\m=dw^\m/ ds$. $o(\ve)$ denotes terms that vanish for $\ve
\ra 0$ in ${\cal S}^\prime$. In the present case these are terms of
the form $\ve^2/R^4$, that converge to zero by power counting. We
have now to perform the limit of this expression for $\ve\ra 0$ in
${\cal S}'$. This analysis can be performed using the techniques
developed in section \ref{extensor}, with the difference that now
one has to keep also terms that are finite for $\ve\ra0$. The
computation is deferred to appendix B where we show that,
 \be\label{riscons}
F^{\n\m}_{\ve} j_{\ve \m}=\left({e\over 4\pi}\right)^2\int\left[
{8\pi\over 3}\left({dw^\n\over ds}+w^2u^\n\right)-{\pi^2\over
2\ve}\left(w^\n-{1\over
4}\,\pa^\n\right)\right]\d^4(x-y(s))\,ds+o(\ve),
 \ee
where $o(\ve)$ denotes again terms that vanish for $\ve \ra 0$ in
${\cal S}^\prime$.

Inserting this in \eref{interm} one sees that the polar
contributions cancel, as they must,  and one obtains \eref{divtem}.

\subsection{Comparison with Rowe's energy--momentum tensor}

The electromagnetic energy--momentum tensor introduced by Rowe
\cite{rowese} is written as the sum of three terms elements of
${\cal S}^\prime$,
 \be \label{6.1}
\Theta^{\m\n}_{em} = \sum^3_{i=1} \Theta^{\m\n}_i,
 \ee
where,
 \ba
\Theta^{\m\n}_1  &=& \partial_\alpha K_1^{\alpha\m\n} +{e^2 \over
16\pi} \int
u^\m \partial^\n \d^4(x - y (s))\,ds, \label{6.2} \\
\Theta^{\m\n}_2  &=& \partial_\alpha K^{\alpha\m\n}_2 -
{e^2\over 6\pi} \int u^\m w^\n \d ^4(x - y(s))\,ds, \label{6.3}\\
\Theta^{\m\n}_3&=& -\left({e\over 4\pi}\right)^2 {\Delta^2\over
(uR)^6}\,R^\m R^\n.
  \label{6.4}
 \ea
The tensors $K_i^{\a\m\n}\in {\cal S}^\prime$ are antisymmetric in
$\a$ and $\m$, and are given by,
 \ba
 K^{\alpha\m\n}_1 &=& {1\over 4}\left({e\over 4\pi}\right)^2 \left(
 \pa^\m\left({R^\a R^\n\over (uR)^4}\right)-\pa^\a\left({R^\m R^\n\over
 (uR)^4}\right)\right), \label{6.5}
\\
K^{\alpha\m\n}_2 &=& \left({e\over 4\pi}\right)^2{1\over (uR)^5}
\left(R^\m \Delta^\a-R^\a\D^\m\right)R^\n. \label{6.6}
 \ea
In these expressions no regularization is needed. Notice, indeed,
that $K^{\alpha\m\n}_2$ contains only ${1/R^2}$ singularities and
that $K^{\alpha\m\n}_1$ contains only derivatives of  $1/R^2$
singularities, so that both tensors are elements of ${\cal S}'$.
This implies that also $\Theta^{\m\n}_1$ and $\Theta^{\m\n}_2$ are
elements of ${\cal S}'$, because the derivative of a distribution is
again a distribution. The terms with the $\d$--functions in
\eref{6.2}, \eref{6.3} are added to ensure the symmetry of each
$\Theta^{\m\n}_i$ separately -- a property that is rather hidden in
Rowe's formulation, and needs to be proven a posteriori. A more
serious practical drawback of the expressions above is that the
derivatives appearing are distributional derivatives, and they can
not be evaluated applying simply the Leibnitz rule, due to the
singularities present on the worldline.

Away from the worldline, i.e. for $x^\m\neq y^\m(\lambda)$, one can
evaluate the derivatives above using the Leibnitz rule, and it is
then straightforward to check that in the complement of the
worldline $\Theta_{em}^{\m\n}$ coincides with the sum of
\eref{t1}--\eref{t3}, with $\ve=0$, and hence with our
$T_{em}^{\m\n}$ in \eref{3.5}.

The comparison of $\Theta_{em}^{\m\n}$ with $T_{em}^{\m\n}$ as
elements of ${\cal S}'$ can be performed applying to the tensors
$K^{\alpha\m\n}_i$ our ${\cal S}'$--regularization,
$K^{\alpha\m\n}_i\ra K^{\alpha\m\n}_{\ve i}$, where the regularized
tensors are given formally again by \eref{6.5}, \eref{6.6}, but with
the replacement $s(x)\ra s_\ve(x)$. On the regularized tensors one
can now use the Leibnitz rule to evaluate $\pa_\a
K^{\alpha\m\n}_{\ve i}$, and perform eventually the ${\cal
S}'$--limit for $\ve\ra 0$. Following this procedure in appendix C
we prove indeed that,
$$
\Theta_{em}^{\m\n}= T_{em}^{\m\n},
$$
as distributions. Here we note only that a comparison of \eref{6.4}
with \eref{t3} yields immediately, since there are only $1/R^2$
singularities present,
 \be \label{6.7} {\cal S}'-\lim_{\ve\ra0} T^{\m\n}_{\ve 3} = \Theta^{\m\n}_3.
 \ee

\vskip1truecm

\section{Proof of variational results}\label{proofaction}

\subsection{Derivation of the Dirac--Lorentz equation}\label{proof2}

In this subsection we proof property 2) of Theorem II, that is
equivalent to the relation \eref{var}. Instead of computing the
variation of $I_\ve$ under a generic variation of $y^\m(s)$, we
compute the variation of the Lagrange density ${\cal L}_\ve$. The
limit of $\d I_\ve$ for $\ve\ra 0$ can then be obtained integrating
$\d{\cal L}_\ve$ between two hypersurfaces, and sending $\ve$ to
zero. Another reason for proceeding in this way is that the explicit
form of $\d{\cal L}_\ve$ will also be used in the proof of Theorem
IV.

In this subsection with ``$\d$" we will always mean a generic smooth
variation w.r.t. $y^\m$. Starting from \eref{3.8} we have then,
 \be\label{vary}
\d {\cal L}_\ve=  -A_{\ve\m} \d j^\m_{\ve} - m_\ve \d\left(\int
\d^{4} (x - y(s))\,ds\right),
 \ee
where, according to what we need in \eref{var}, we have replaced --
after variation -- $A_\m\ra A_{\ve\m}$. While the variation of the
second term is trivial, to vary the first term it is more convenient
to write,
 \be \label{5.4}
-A_{\ve\m} \d j^\m_{\ve} = -\d\left(A_{\ve\m} j^\m_{\ve}\right) + \d
A_{\ve\m}j^\m_\ve,
 \ee
since $\d A_{\ve\m}$ is simpler then $\d j^\m_{\ve}$. The variation
of the first term in this expression can be obtained from the ${\cal
S}'$--expansion,
 \be
j^\m_{\ve} A^\n_{\ve} = \left({e\over 4\pi}\right)^2 \int \left(
{3\pi^2\over 4\ve} \,u^\m u^\n - 4\pi \,u^\m w^\n +o(\ve)
\right)\d^4(x-y(s))\,ds, \label{5.5}
 \ee
upon contracting $\m$ and $\n$. This expansion can be derived using
the method illustrated in appendix B, see \eref{esempio}.

The evaluation of the second term in \eref{5.4} is more cumbersome,
and the calculation is deferred to appendix D. The result is,
 \ba \d A_{\ve\m}j^\m_\ve&=& \left({e\over 4\pi}\right)^2\left[\int
\left(-{3 \pi^2\over 8\ve} (w\,\d y)- {8 \pi \over 3} \left(
{dw^\m\over ds}+w^2 u^\m\right)\d y_\m\right)\d^4(x-y(s))\,ds
\right.\nn
\\
&&\left.+ \pa_\m \int \left(4\pi (w\,\d y)u^\m-{\pi^2\over 8\ve}(\d
y^\m- (u\,\d y )u^\m)\right)\d^4(x-y(s))\,ds\right] + o(\ve).
\label{varj}
 \ea
Using this in \eref{5.4} one obtains eventually,
 \ba
 \d {\cal L}_\ve&=& \int
\left(m w^\m- {e^2 \over 6\pi} \left( {dw^\m\over ds}+w^2 u^\m
\right)\right)\d y_\m\,\d^4(x-y(s))\,ds+ \label{generale}
\\
&&+ \pa_\m \int \left({e^2\over4\pi}(w\,\d
y)\,u^\m+\left(m+{e^2\over 64\ve}\right)(\d y^\m- (u\,\d y
)u^\m)\right)\d^4(x-y(s))\,ds + o(\ve),\nn
 \ea
which is our final result, holding for generic variations of $y$.
The first line is finite as $\ve\ra 0$, and the second line, which
corresponds to a four--divergence, contains a polar term in $\ve$.

To derive the equations of motion we compute $\d I_\ve$ integrating
$\d {\cal L}_\ve$ between two hypersurfaces, and impose that on them
$\d y^\m(s)$ vanishes. The four--divergence drops then out --
together with the polar term -- and the first line  \eref{generale}
gives \eref{var}. This concludes the proof of Theorem II.

\subsection{Proof of Theorem III}\label{proof3}

The evaluation of the regularized Lagrangian on the solution
$A^\m=A^\m_\ve$ requires to evaluate, using \eref{2.18} and
\eref{lim3},
 \ba
 -{1\over 4} F_{\ve \m\n} F_{\ve}^{\m\n} &=& {1\over 2}\left({e\over
4\pi}\right)^2  \left({1\over (uR)^4} - {\ve^2\over
(uR)^6} +o(\ve) \right)\nn\\
&=&{1\over 2}\left({e\over 4\pi}\right)^2  \left({1\over (uR)^4} -
{\pi^2\over 4\ve} \int \d^4(x-y(s))\,ds +o(\ve)\right).\nn
 \ea
This gives, using \eref{5.5},
 \ba
 {\cal L}_{\ve}(A_\ve,y)&=&-{1\over 4} F_{\ve \m\n} F_{\ve}^{\m\n} -A_{\ve\m}
 j^\m_{\ve}-m_\ve\int \d^4(x-y(s))\,ds\nn\\
&=& {1\over 2}\left({e\over 4\pi}\right)^2  \left({1\over (uR)^4} -
{\pi^2\over \ve}\int \d^4(x-y(s))\,ds\ \right)-m \int
\d^4(x-y(s))\,ds+ o(\ve).\nn
 \ea
Applying to a test function one gets,
 \ba\nn
\lim_{\ve\ra 0}{\cal L}_{\ve}(A_\ve,y)(\vp)&= &\lim_{\ve\ra
0}{1\over 2}\left({e\over 4\pi}\right)^2 \int ds \int
{2\d_+(x^2-\ve^2)\over (ux)^3}[\vp(x+y(s))-\vp(y(s))]\,d^4x\\
&&-m\int \vp(y(s))\,ds,\nn
 \ea
where we used,
$$
\int d^4x \,{2\d_+(x^2-\ve^2)\over (ux)^3}= {\pi^2\over \ve}.
$$
The integral above converges now for $\ve\ra 0$, and we can set
$\ve=0$ obtaining the conditionally convergent integral,
$$
\lim_{\ve\ra 0}{\cal L}_{\ve}(A_\ve,y)(\vp) ={1\over 2}\left({e\over
4\pi}\right)^2 \int ds \int {2\d_+(x^2)\over
(ux)^3}[\vp(x+y(s))-\vp(y(s))]\,d^4x -m\int \vp(y(s))\,ds.
$$
Shifting $x\ra x-y(s)$ and integrating out the $\d$--function one
obtains \eref{lagren}.

\subsection{Proof of the Noether Theorem IV}\label{proof4}

The translation invariance of the regularized Lagrangian is
expressed by  the identity \eref{deltal}, $ \d_t {\cal L}_\ve=0$,
where ${\cal L}_\ve$ is given in \eref{3.8}, and $\d y^\m = a^\m, $
$\d A^\m = a^\n\pa_\n A^\m $.

After standard steps one obtains,
 $$ \d_t {\cal L}_\ve= (\pa_\m
F^{\m\n}-j^\n_\ve)\d A_\n-\pa_\m(\d A_\n F^{\m\n}) +A_\m \d_y
j^\m_\ve+\d_y \left(m_\ve \int \d^4(x-y(s))\,ds\right)-a^\m \pa_\m
{\cal L}_\ve,
 $$
where $\d_y$ means variation w.r.t. $y^\m$. We evaluate this
expression now on the solutions of the regularized Maxwell equation,
$A^\m=A^\m_\ve$, and on those of the Lorentz--Dirac equation. The
first term goes then to zero, and for the combination,
$$
A_{\ve\m} \d_y j^\m_\ve+\d_y \left(m_\ve \int
\d^4(x-y(s))\,ds\right),
$$
we can use the general expansion \eref{vary}, \eref{generale}, with
$\d y^\m=a^\m$, with opposite sign. Thanks to the Lorentz--Dirac
equation, the expression \eref{generale} reduces to a
four--divergence and hence $\d_t {\cal L}_\ve$ becomes a
four--divergence, too. Using again \eref{5.5} to evaluate
$A_{\ve\n}j^\n_\ve$ appearing in $\pa_\m{\cal L}_\ve$, after simple
algebra one obtains,
 \be\label{5.10}
\d_t {\cal L}_\ve=a_\n\,\partial_\m \widehat T^{\m\n}_\ve
+o(\ve)=0, \ee where,
 \ba
 \widehat T^{\m\n}_\ve &\equiv& F^{\rho\m}_{\ve}\pa^\n
A_{\ve\rho}+ {1\over 4} \,\eta^{\m\n} F^{\rho\sigma}_{\ve}
F_{\ve\rho\sigma} + m\int u^\m u^\n \d^4(x-y(s))\,ds\nn\\
&&+
 \left({e\over 4\pi}\right)^2 \int
 \left({\pi^2\over 4\ve}\,u^\m u^\n - 4\pi u^\m w^\n +
{\pi^2\over 8\ve}\,\eta^{\m\n}\right)\d^4(x - y(s)) +o(\ve).
 \label{5.11}
 \ea
Even if it is not obvious from the derivation -- see however the
footnote below -- it can be seen hat the limit,
 \be\label{tcanon}
{\cal S}'-\lim_{\ve\ra 0}\widehat T^{\m\n}_\ve  \equiv \widehat
T^{\m\n},
 \ee
exists. \eref{5.10} implies then that the energy--momentum tensor
$\widehat T^{\m\n}$ is conserved. It is immediately seen, however,
that this tensor does not coincide with the one constructed in
Theorem I, see \eref{3.4}, nor is it symmetric in $\m$ and $\n$. On
the other hand this was to be expected, since the Noether theorem
leads in general to the {\it canonical} energy--momentum tensor, not
to the symmetric one.

From the general theory we know however that, if the Lagrangian
density is not only translation invariant but also Lorentz
invariant, there exists always a symmetrization procedure for the
canonical tensor. In the case of classical Electrodynamics this
procedure requires to add the divergenceless term $-\partial_\rho
(F^{\rho\m} A^\n)$. In the case at hand this suggests then to add
the divergenceless term \footnote{The product $F^{\rho\m}_\ve
A^\n_\ve$ contains terms that behave as $1/R^3$, if $\ve =0$. This
means that a priori its ${\cal S}'$--limit for $\ve\ra 0$ contains
logarithmic divergences, $\sim \ln\ve$. However, by inspection i.e.
applying the product to a test--function and taking $\ve\ra 0$ one
checks that these divergences cancel out, and the limit exists. This
means that also,
$$
{\cal S}'-\lim_{\ve\ra 0} \pa_\rho (F^{\rho\m}_\ve A^\n_\ve),
$$
exists. This fact, together with \eref{5.13}, provides an indirect
proof for the existence of the limit \eref{tcanon}.}, see
\eref{5.5},
 \ba
 \label{symm}
 -\pa_\rho (F^{\rho\m}_\ve A^\n_\ve)&=& - F^{\rho\m}_\ve \pa_\rho A^\n_\ve -j^\m_\ve A^\n_\ve
 \nn\\
&=&-F^{\rho\m}_\ve \pa_\rho A^\n_\ve +\left({e\over 4\pi}\right)^2
\int \left( -{3\pi^2\over 4\ve} \,u^\m u^\n + 4\pi \,u^\m w^\n
+o(\ve) \right)\d^4(x-y(s))\,ds.\nn
 \ea
Adding this to \eref{5.11} one sees that eventually one gets a
symmetric tensor, being indeed,
 \be \label{5.13}
 \widehat T^{\m\n}_\ve - \partial_\rho
(F^{\rho\m}_{\ve} A^\n_{\ve})= T^{\m\n} +o(\ve),
 \ee
where $T^{\m\n}$ is defined in \eref{3.4}. This concludes the proof
of the Noether theorem.

\bigskip

\paragraph{Acknowledgements.}
This work is supported in part by the European Community's Human
Potential Programme under contract MRTN-CT-2004-005104,
``Constituents, Fundamental Forces and Symmetries of the Universe".

\vskip2truecm

\section{Appendix A: Invariant integrals}

In the proofs throughout the paper one encounters the  invariant
integrals,
 \ba
I_\m&=&\int d^4x \,{2\d_+(x^2-1)\over (ux)^N}\, x_\m\nn\\
I_{\m\n}&=&\int d^4x \,{2\d_+(x^2-1)\over (ux)^N}\,x_\m x_\n,\nn\\
I_{\m\n\r}&=&\int d^4x \,{2\d_+(x^2-1)\over (ux)^N}\, x_\m x_\n\nn
x_\r,
 \ea
where $\d_+(x^2-1)\equiv H(x^0)\d(x^2-1)$. These integrals can be
evaluated considering $u^\m$ as an unconstrained variable, not
satisfying $u^2=1$, and writing, for example,
$$
I_{\m\n}= {1\over (N-1)(N-2)}{\pa\over \pa u^\m} {\pa\over \pa u^\n}
\int d^4x \,{2\d_+(x^2-1)\over (ux)^{N-2}}.
$$
This reduces  these integrals to the calculation of the ``generating
function",
$$
\int d^4x \,{2\d_+(x^2-1)\over (ux)^N}= {\pi^{3/2}\over (u^2)^{N/2}}
 {\Gamma\left({N\over 2}-1\right)\over \Gamma\left({N+1\over
 2}\right)}.
$$
After taking the derivatives w.r.t. $u^\m$ one sets again $u^2=1$.
For example,
 $$
\int d^4x \,{2\d_+(x^2-1)\over (ux)^3}= {\pi^2\over (u^2)^{3/2}},
 $$
and hence,
$$
\int d^4x \,{2\d_+(x^2-1)\over (ux)^5}\,x_\m x_\n= {1\over 3\cdot
4}{\pa\over \pa u^\m} {\pa\over \pa u^\n} \int d^4x
\,{2\d_+(x^2-1)\over (ux)^3}={\pi^2\over 4(u^2)^{7/2}}\left(5u^\m
u^\n-u^2\eta^{\m\n}\right).
$$
Setting $u^2=1$ this gives \eref{invint}.

\vskip1truecm

\section{Appendix B: proof of \eref{riscons}}

Before taking the limit of $\ve \ra 0$ it is convenient to rewrite
\eref{4.14} in the form,
 \be
F^{\n\m}_{\ve} j_{\ve \m}= \left({e\over 4\pi}\right)^2 \ve^2 \left(
 \left({2w^2 \over (uR)^6} -{(b R)\over (uR)^7}\right)R^\n -{3w^\n\over
 (uR)^6}
+{1\over 2}((wR)-1)^2\pa^\n{1\over (uR)^6}\right) +o(\ve).
 \label{4.14bis}
 \ee
We have used,
$$
\pa^\n (uR)=u^\n +((wR)-1) \,{R^\n\over (uR)},
$$
and we have omitted the term,
$$
3\ve^2 {(wR)w^\n\over (uR)^6},
$$
because it goes to zero for $\ve\ra0$ in ${\cal S}'$, for
kinematical reasons. More precisely, after applying it to a test
function and integrating over $d^4x$, for invariance reasons the
factor $R^\m$ in the numerator gets replaced by $u^\m$, and
$(uw)=0$. This will become clear below. Regarding the last term in
\eref{4.14bis} it is convenient to note the identity,
 \ba
{\ve^2\over 2}((wR)-1)^2\pa^\n{1\over (uR)^6}&=& {\ve^2\over 2}\,
\pa^\n \left[{((wR)-1)^2\over(uR)^6}\right] -\ve^2 {((wR)-1)\over
(uR)^6}\, \pa^\n(wR) \nonumber\\
&=& {\ve^2\over 2} \,\pa^\n {1\over(uR)^6}+\ve^2{(bR)\over (u
R)^7}\,R^\n +\ve^2{w^\n\over (uR)^6} + o(\ve),
 \ea
where the $o(\ve)$ terms are in the same sense as above. Inserting
this in \eref{4.14bis} one sees that the $(bR)$--term cancels
getting,
 \be
 \label{4.16}
F^{\n\m}_{\ve} j_{\ve\m} =\left({e\over 4\pi}\right)^2 \ve^2 \left(
{2w^2 \over (uR)^6}\, R^\n - {2 w^\n\over (u R)^6} + {1\over 2}\,
\pa^\n{1\over (u R)^6}\right) +o(\ve).
 \ee
The behaviour of the three terms present, for $\ve \ra 0$ in ${\cal
S}'$, are as follows,
 \ba
{2\ve^2 w^2\over (u R)^6}\, R^\n & =&{8\pi\over 3} \int w^2
u^\n\d^4(x - y (s))\,ds +o(\ve),\label{lim1}\\
-{2\ve^2 w^\n\over (uR)^6}&=&\int\left({8\pi\over 3} {dw^\nu\over
ds} -{\pi^2\over 2\ve}w^\n\right)\d^4(x - y (s))\,ds +o(\ve),\label{lim2}\\
{\ve^2\over 2(uR)^6}&=&{\pi^2\over 8\ve} \int \d^4(x - y (s))\,ds
+o(\ve).\label{lim3}
 \ea
For the proofs of these relations, that require to apply the l.h.s.
to a test function and to analyze the limit $\ve \ra 0$, one can use
the technicalities developed in section \ref{extensor}. We report
explicitly the proof of \eref{lim2},
 \ba
 \left(-{2\ve^2 w^\n\over (uR)^6}\right)\,(\vp)&=&
-2\ve^2 \int ds \int d^4x{ 2 \d_+ (x^2 -
\ve^2)\over (ux)^5} \,w^\n\vp(y(s)+ x)\nn\\
&=& -{2\over \ve} \int ds \int d^4x{ 2 \d_+ (x^2 - 1)\over (ux)^5}
\,w^\n\vp(y(s)+\ve x)\nn\\
&=& -2 \int ds \int d^4x{2\d_+ (x^2 - 1)\over (ux)^5}\,w^\n\,
\left({1\over \ve}\,\vp
(y(s)) +  x^\alpha \pa_\alpha \vp(y(s)) + o(\ve)\right)\nn\\
&=& -2 \int ds \left({\pi^2\over 4\ve}\,w^\n \vp(y(s)) +{4\pi\over
3}\,
w^\n u^\alpha \pa_\alpha \vp (y(s)) + o(\ve)\right)\nn\\
&=& \int ds \left(-{\pi^2\over 2\ve} w^\n + {8\pi\over 3}
{dw^\n\over ds}+o(\ve)\right) \vp (y(s)),\label{esempio}
 \ea
where we used the invariant integrals of appendix A.

Inserting \eref{lim1}--\eref{lim3} in \eref{4.16}, one obtains
\eref{riscons}.

\vskip1truecm

\section{Appendix C: Rowe's energy--momentum tensor}\label{appa}

Before comparing $T^{\m\n}_{em}$ of \eref{3.5} with
$\Theta^{\m\n}_{em}$ in \eref{6.1}, we cast the former in a slightly
different form writing, see \eref{t1}--\eref{t3},
 \be\label{newsplit}
T^{\m\n}_{em}=  {\cal S}'-\lim_{\ve\ra0} \left(D^{\m\n}_{\ve 1}+
D^{\m\n}_{\ve 2}+T^{\m\n}_{\ve 3}\right),
 \ee
where,
 \ba
D^{\m\n}_{\ve 1}&=&T^{\m\n}_{\ve1}  - 2\left({e\over 4\pi}\right)^2
{(wR)\over (uR)^6}\,(R^\m R^\n- {\rm tr})- {e^2\over 32\,\ve} \int
\left(u^\m u^\n - {1\over 4} \,\eta^{\m\n}\right)\d^4
(x - y (s))\,ds,\nn\\
D^{\m\n}_{\ve 2}&=&T^{\m\n}_{\ve2}  + 2\left({e\over 4\pi}\right)^2
{(wR)\over (uR)^6}\,(R^\m R^\n- {\rm tr}),\nn
 \ea
The advantage of this form of presenting $T^{\m\n}_{em}$ is that the
tensors $D^{\m\n}_{\ve i}$ converge {\it separately} for $\ve \ra 0$
in the topology of ${\cal S}'$, as does $T^{\m\n}_{\ve 3}$. We have
added and subtracted the same term from $ T^{\m\n}_{\ve 1}$ and
$T^{\m\n}_{\ve 2}$, to eliminate the logarithmic divergence from
both. This can be seen from \eref{div1}--\eref{div3}, noting that,
$$
\left.{(wR)\over (uR)^6}\,R^\m R^\n\right|_{div}= {8\pi\over
3}\,\ln\ve\int u^{(\m} w^{\n)}\d^4(x-y(s))\,ds.
$$

We evaluate now the four--divergencies appearing in \eref{6.2},
\eref{6.3}, using the regularized versions $K_{\ve i}^{\alpha\m\n}$
for \eref{6.5}, \eref{6.6}, obtained replacing $s(x)\ra s_\ve(x)$.
Then one can apply Leibnitz' rule and, taking into account that now
$R^2=\ve^2\neq 0$, after a straightforward computation one obtains,
 \ba
\pa_\a K_{\ve1}^{\a\m\n}& =&D^{\m\n}_{\ve1} + {e^2\over 32\,\ve}
\int \left(u^\m u^\n - {1\over 4} \,\eta^{\m\n}\right)\d^4 (x - y
(s))\,ds\label{a1}\\
&&+ \left({e\over 4\pi}\right)^2 {\ve^2\over 4}
 \left[15{(wR)-1\over(u R)^7} \,u^\m R^\n
 + {7u^\m u^\n\over (uR)^6} -{3w^\m R^\n\over (uR)^6}  + {2- 6 (w R)\over (uR)^6}\,
\eta ^{\m\n}\right], \label{a2}\\
\pa_\a
K_{\ve2}^{\a\m\n}&=&D^{\m\n}_{\ve2}+\left({e\over4\pi}\right)^2\ve^2
\left[5 {(wR)-1\over (uR)^7}\D^\m R^\n+{w^\m R^\n\over (uR)^6} +
{1\over (uR)^6}\left((bR)u^\m-(uR)b^\m\right)R^\n\right.\nn
\\
&&\left.+ {1\over (uR)^6}(2\D^\m u^\n+\D^\n u^\m)+{3(wR)\over
2(uR)^6}\eta^{\m\n}\right].\label{a4}
 \ea
In $\pa_\a K_{\ve1}^{\a\m\n}$ we have put in evidence the polar
part, proportional to $u^\m u^\n - {1\over 4} \,\eta^{\m\n}$, and in
$\pa_\a K_{\ve1}^{\a\m\n}$ we have set $b^\m=dw^\m/ds$.

We can now consider the ${\cal S}'$--limit for $\ve\ra 0$ of these
relations, applying them to a test--function and sending $\ve\ra0$.
The calculations, that are straightforward but a bit lengthy, can be
performed with the techniques illustrated in appendix B, see
\eref{esempio}. Since the limits of $\pa_\a K_{\ve i}^{\a\m\n}$ for
$\ve\ra0$ exist by construction, the r.h.s. of the above relations
must also admit finite limits. It can indeed be seen that the square
bracket in \eref{a2} exhibits a divergent contribution that cancels
exactly the polar term in \eref{a1}, while the square bracket in
\eref{a4} admits a finite limit by power counting. The results are,
 \ba
\pa_\a K_1^{\a\m\n}&=& \left({\cal S}'-\lim_{\ve\ra0} D^{\m\n}_{\ve
1}\right) + {e^2\over \pi} \int\left({1\over 6}(u^\m w^\n+u^\n
w^\m)-{1\over 16}\,u^\m \pa^\n\right)\d^4(x-y(s))\,ds,\nn
\\
\pa_\a K_2^{\a\m\n}&=& \left({\cal S}'-\lim_{\ve\ra0} D^{\m\n}_{\ve
2}\right) - {e^2\over 6\pi} \int u^\n w^\m \d^4(x-y(s))\,ds.\nn
 \ea
Inserting these expressions in \eref{6.2}, \eref{6.3} one gets the
symmetric and traceless tensors,
 \ba
\Theta^{\m\n}_1 &=&\left({\cal S}'-\lim_{\ve\ra0} D^{\m\n}_{\ve
1}\right)  + {e^2\over 6\pi} \int (u^\m w^\n+u^\n w^\m) \d^4(x -
y(s))\,ds,\\
\Theta^{\m\n}_2 &=&\left({\cal S}'-\lim_{\ve\ra0} D^{\m\n}_{\ve
2}\right)  - {e^2\over 6\pi} \int (u^\m w^\n+u^\n w^\m) \d^4(x -
y(s))\,ds.
 \ea
These formulae, together with \eref{6.7} and \eref{newsplit}, imply
then $\Theta^{\m\n}_{em}=T^{\m\n}_{em}$, q.e.d.

\vskip1truecm

\section{Appendix D: Proof of \eref{varj}}\label{appb}

To evaluation of $\d A_{\ve\m}j^\m_\ve $ requires first to evaluate
$\d A_{\ve\m}$ under a generic smooth variation of $y^\m$. This
evaluation is complicated by the fact that $A_\ve^\m$  -- see
formula \eref{2.8} -- depends on $y^\m$ explicitly,  but also
implicitly through the (regularized) retarded proper time function
$s_\ve(x)$, defined in \eref{2.15}. To determine  $\d A_{\ve\m}$ it
is convenient to use the following shortcut.

We compute $\d A_\ve^\m$ applied to a test function $\vp\in {\cal
S}$, inserting a $\d$--function as in \eref{4.3}, and then shifting
$x\ra x+y(s)$,
 \ba
\int\d A^\m_{\ve}\,\vp\, d^4x &=& \d \int A^\m_{\ve}\,\vp \,d^4x ={e
\over 4 \pi}\,\d \int ds \int d^4 x
\,2\d_+(x^2-\ve^2) u^\m(s) \vp(x+y(s))\nn\\
&=& {e \over 4 \pi} \int ds \int d^4 x \,2\d_+(x^2-\ve^2)\left(
{d\,\d
y^\m\over ds} \,\vp(x+y(s)) +u^\m\d y^\n\pa_\n  \vp(x+y(s))\right)\nn\\
 &=&
{e \over 4 \pi} \int d^4 x\left[{1\over (uR)}\,{d\d y^\m\over ds}
-\pa_\n\left({u^\m\d y^\n \over (uR)}\right)\right]\vp,\nn
 \ea
where in the square bracket all kinematical quantities are evaluated
at $s=s_\ve(x)$. This gives,
 \ba
\d A^\m_{\ve}&=&{e \over 4 \pi}\left({1\over (uR)}\,{d\d y^\m\over
ds}
-\pa_\n\left({u^\m\d y^\n \over (uR)}\right)\right)\nn\\
 &=&
{e \over 4 \pi}\left({\ell'^\m\over (uR)}-{(R\ell\,')\over
(uR)^2}u^\m-{(R\ell)\over (uR)^3}(\D^\m+u^\m)\right),  \label{7.1}
 \ea
where,
$$
\ell^\m\equiv \d y^\m-(u\d y)u^\m, \quad (u\ell)=0,   \quad
\ell'^\m\equiv{d \ell^\m \over ds}.
$$
The appearance of the combination $\ell^\m$ is due to the invariance
of $A^\m_\ve$ under reparametrization of the worldline, i.e. $\d
A^\m_\ve=0$ if $\d y^\m=\d\lambda \,u^\m $.

Multiplying out \eref{7.1} and \eref{curreg} one obtains then,
 \ba\label{varfin}\nn
\d A_{\ve\m}j^\m_\ve &=&\left({e \over 4 \pi}\right)^2 \ve^2 \left[
{(bR)-(uR)(ub)\over (uR)^7}\,(R\ell)+ 3(1-(wR))\cdot\right.
\\
&&\cdot\left.\left({((u\ell')+(\D\ell'))(uR)-(R\ell')(1-(wR))\over
(uR)^7}- {(1-(wR))^2+(uR)^2w^2\over
(uR)^8}\,(R\ell)\right)\right]\nn
\\&&+o(\ve),
 \ea
where the terms $o(\ve)$ correspond to expressions of the type
$\ve^2/R^4$, that converge to 0 in the topology of ${\cal S}'$, by
power counting. To evaluate the limit of this expression under
$\ve\ra0$ in ${\cal S}'$, one proceeds precisely as in
\eref{esempio}. In this way the evaluation of the limit of all terms
above is reduced to the determination of the invariant integrals of
appendix A.

Lets consider e.g. in the expression above the most singular term,
corresponding to $\ve^2 (R\ell)/(uR)^8$, that by power counting
would give rise to a double pole $1/\ve^2$ in that,
 \ba
\left(\ve^2 {(R\ell)\over (uR)^8}\right)(\vp) &=& \ve^2 \int ds \int
d^4x{ 2 \d_+ (x^2 -
\ve^2)\over (ux)^7} \, (x\ell)\vp(y(s)+ x)\nn\\
&=& {1\over \ve^2} \int ds \int d^4x{ 2 \d_+ (x^2 - 1)\over (ux)^7}
\,(x\ell)  \vp(y(s)+\ve x)\nn\\
&=& \int ds \int d^4x{2\d_+ (x^2 - 1)\over (ux)^7}\,(x\ell)
\left[{1\over \ve^2}\,\vp (y(s)) +  {1\over \ve} x^\alpha \pa_\alpha
\vp(y(s))\right.\nn\\
&&\hskip5truecm+{1\over 2} \left.x^\a x^\b \pa_\a\pa_\b\vp(y(s)) +
o(\ve)\right].\nn
 \ea
We compute the invariant integrals according to appendix A,
 \ba
\int d^4x{2\d_+ (x^2 - 1)\over (ux)^7}\,x^\m&=&{8\pi\over 15}u^\m,\nn\\
\int d^4x{2\d_+ (x^2 - 1)\over (ux)^7}\,x^\m x^\a &=&{\pi^2\over 24}
\left(7u^\m u^\a-\eta^{\m\a}\right),\nn\\
\int d^4x{2\d_+ (x^2 - 1)\over (ux)^7}\,x^\m x^\a x^\b&=&{4\pi\over
15} \left(8u^\m u^\a
u^\b-\eta^{\m\a}u^\b-\eta^{\a\b}u^\m-\eta^{\b\m}u^\a\right). \nn
 \ea
Since $(u\ell)=0$ we see that the double pole cancels, and the
remaining two terms give a $1/\ve$ contribution and a finite part.
After integration by parts one obtains,
$$
\left(\ve^2 {(R\ell)\over (uR)^8}\right)(\vp)=\int \left({4\pi\over
15}\,\ell'^\m -{\pi^2\over 24
\ve}\,\ell^\m\right)\pa_\m\vp(y(s))\,ds +o(\ve),
$$
i.e. in ${\cal S}'$,
$$
\left(\ve^2 {(R\ell)\over (uR)^8}\right)=-\pa_\m \int
 \left({4\pi\over
15}\,\ell'^\m -{\pi^2\over 24 \ve}\,\ell^\m\right) \d^4(x-y(s))\,ds
+ o(\ve).
$$
All other terms in \eref{varfin} by power counting give simple pole
contributions and finite parts, too, and a straightforward but a bit
lengthy calculation gives \eref{varj}.

\end{document}